\documentclass[journal]{IEEEtran}
\ifCLASSINFOpdf
\else
\fi
\UseRawInputEncoding
\usepackage{cite}
\usepackage{amsmath,amssymb,amsfonts}
\usepackage{algorithmic}
\usepackage{graphicx}
\usepackage{textcomp}
\usepackage{xcolor}
\usepackage{mathrsfs}
\usepackage{algorithm}
\usepackage{eqparbox}
\usepackage{caption}
\usepackage{graphicx}
\usepackage{float}
\usepackage{epstopdf}
\usepackage{subfigure}

\newtheorem{theorem}{\bf Theorem}  [section]
\newtheorem{corollary}{\bf Corollary}  [section]

\newtheorem{definition}{\bf Definition}[section]

\def\BibTeX{{\rm B\kern-.05em{\sc i\kern-.025em b}\kern-.08em
    T\kern-.1667em\lower.7ex\hbox{E}\kern-.125emX}}
\begin{document}
\title{Active Defense Analysis of Blockchain Forking through the Spatial-Temporal Lens}
\author{Shengling~Wang,~\IEEEmembership{Senior Member,~IEEE,}
        Ying~Wang,
        Hongwei~Shi,
        and Qin~Hu
\thanks{Shengling Wang is with the School of Artificial Intelligence and the Faculty of Education, Beijing Normal University, Beijing, China. E-mail: wangshengling@bnu.edu.cn.}
\thanks{Ying Wang is with the School of Artificial Intelligence, Beijing Normal University, Beijing, China. E-mail: 201921210035@mail.bnu.edu.cn.}
\thanks{Hongwei Shi (Corresponding author) is with the School of Artificial Intelligence, Beijing Normal University, Beijing, China. E-mail: hongweishi@mail.bnu.edu.cn.}
\thanks{Qin Hu is with the Department of Computer and Information Science, Indiana University - Purdue University Indianapolis, IN, USA.E-mail:qinhu@iu.edu.}

}

\markboth{Journal of \LaTeX\ Class Files,~Vol.~14, No.~8, August~2015}%
{Shell \MakeLowercase{\textit{et al.}}: Bare Demo of IEEEtran.cls for IEEE Journals}
\maketitle

\begin{abstract}
Forking breaches the security and performance of blockchain as it is symptomatic of distributed consensus, spurring wide interest in analyzing and resolving it.  The state-of-the-art works can be categorized into two kinds: experiment-based and model-based. However, the former falls short in {\it exclusiveness} since the derived observations are scenario-specific. Hence, it is problematic to abstractly reveal the crystal-clear forking laws. Besides, the models established in the latter are {\it spatiality-free}, which totally overlook the fact that forking is essentially an undesirable result under a given topology. Moreover, few of the ongoing studies have yielded to the active defense mechanisms but only recognized forking passively, which impedes forking prevention and cannot deter it at the source. In this paper, we fill the gap by carrying out the active defense analysis of blockchain forking from the spatial-temporal dimension. Our work is featured by the following two traits: 1) {\it dual dimensions}. We consider the spatiality of blockchain overlay network besides temporal characteristics, based on which, a spatial-temporal model for information propagation in blockchain is proposed; 2) {\it active defense}. We hint that shrinking the {\it long-range link factor}, which indicates the remote connection ability of a link, can cut down forking completely fundamentally. To the best of our knowledge, we are the first to inspect forking from the spatial-temporal perspective, so as to present countermeasures proactively. Solid theoretical derivations and extensive simulations are conducted to justify the validity and effectiveness of our analysis.
\end{abstract}

\begin{IEEEkeywords}
Blockchain, forking, spatial-temporal dimension, active defense.
\end{IEEEkeywords}
\IEEEpeerreviewmaketitle
\section{Introduction}

\IEEEPARstart{B}{lockchain} has been burgeoning in a plethora of domains since its inception \cite{bitcoin,ETH,life,ghosh}, due to the prominent traits of decentralization, transparency, traceability and immutability. Essentially, blockchain is a linear chain documenting valid transactions in a form of blocks, with each being approved by the distributed participants through {\it consensus}.  The consensus mechanism functions nontrivially in preserving blockchain security since it guarantees that all the honest participants (or nodes) curate the unique shared chain, or denoted as the {\it main chain}, so as to maintain data consistency. However, there may arise disagreements among nodes over which the main chain is, and when this happens, we call the blockchain is {\it forking}.

\begin{figure}
  \centering
  \includegraphics[width=6cm]{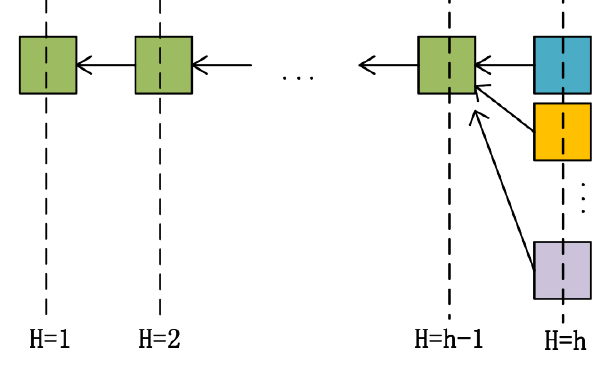}\\
  \caption{The illustration of forking. The green blocks are on the main chain, while at height $H=h$, there are more than 1 block, that is $\mathcal{B}_h=\{\mathcal{B}_{blue},\mathcal{B}_{orange},...,\mathcal{B}_{purple}\}.$}\label{fig:forking}
\end{figure}

Technically, forking refers to the phenomenon that there are a set of blocks $\mathcal{B}_h$ at the same height $h$ at a time, i.e., $|\mathcal{B}_h|>1$ \cite{black}, as shown in Fig. \ref{fig:forking}. This suggests  divergent blockchains are maintained among nodes, leading decentralized consensus to fail. Forking can be {\it unintentional} or {\it intentional} depending on the motivations behind. By ``unintentional", we mean forking is aroused due to the inefficient propagation of the overlay network without any adversary, while the ``intentional" forking involves malicious attackers who aim at inserting new features for security breaches. It is, therefore, crucial to analyze forking since it destroys the distributed trust, threatening the security and performance of blockchain.

The state-of-the-art works on forking can be classified into two kinds, namely the experiment-based \cite{chen,proximity,zhang} and the model-based \cite{P2P,Xiao,corking,impact}. As the names suggest, the former studies clarify and evaluate the insights via experimental simulations while those of the latter are through mathematical deductions. However, the experiment-based analyses are {\it exclusive}, which means the observations drawn from one study may not be feasible for another since they are scenario-specific. This makes the experiment-based analysis an ill-posed problem in extracting decisive forking laws from the top level, falling short in guiding countermeasures effectively.

To remedy the deficiency of the experiment-based studies, the model-based analyses heaved in sight. However, most of them are constrained to the representation of time dimension, without considering  the spatial characteristics of blockchain topology. Nonetheless, forking is essentially an undesirable result of main chain propagation dynamics, which is inevitably bound to the exact positions and connectivities of nodes located in the network. Consequently, such {\it spatiality-free} forking models are incompetent to figure out the fundamental mechanism of forking. 
Moreover, few studies have yielded to the {\it active defense} mechanisms for thwarting forking but only recognized it passively, which caps forking prevention and cannot deter it from the root, thus motivating our work. 

We face the {\it heterogeneity-in-design} challenge in realizing an active defense analysis of blockchain forking through the spatial-temporal lens. That is, a fine-grained quantitative analysis with spatiality requires us to {\it embed the connection heterogeneity into the main chain propagation mechanism.} To resolve this, we endow ``expansivity" for connections by differentiating them as the {\it short-range link} and the {\it long-range link}. Specifically, short-range links exist among direct neighbors while the long-range link can connect remote nodes probabilistically. As such, the short-range links have no expansivity since they can only spread information among geographically adjacent nodes, while the long-range links can transmit over a long distance with various probabilities, so that we claim they have greater expansivity. Accordingly, it is the differences in expansivity that mirror the connection heterogeneity. In addition, we take advantage of the two-dimensional grid network in \cite{small} to abstractly model the blockchain overlay network since recent theoretical developments have stated blockchain as a {\it small-world phenomenon} \cite{tnse,tao,ma}. In doing so, we come up with the definition of {\it layer propagation} conceptually which characterizes the spread of information as the {\it ripple effect}.

Despite the benefits, the introduction of short-/long- range links increases the complexity of {\it spatial-temporal coupling analysis}, which is referred to the second challenge in our paper. Detailedly, the remote transfer ability of the long-range links may accelerate main chain propagation with great uncertainty, complicating the propagation dynamics as a consequence. To deal with this, we thoroughly inspect the probability distributions of main chain propagation triggered by short-range links and long-range links respectively and propose two concepts, i.e., the {\it activation time} and {\it activation degree}, for modeling purposes, where the former denotes the time when the nodes of each layer first receive the main chain while the latter indicates the number of nodes that have received the main chain at any time. These two terms provide a direct clue for further depicting unintentional forking probability and intentional forking robust level, laying foundations for presenting an active defense mechanism to fight against forking.

To conclude, the main contributions of our paper can be summarized as follows:
\begin{itemize}
  \item \textbf{Spatial-temporal propagation model of the main chain.} To begin with, we propose the first spatial-temporal model for the main chain propagation, by specifying the blockchain network as a two-dimensional grid with short-/long- range links. In light of this, the forking probability and robust level in unintentional and intentional scenarios are proposed, where the former is analyzed in various time scales, facilitating countermeasures with different intensities.
  \item \textbf{Active defense mechanism proposal.} Our work concludes that dwindling the {\it long-range link factor} to change the connectivity of blockchain topology serves to cut down forking completely from the root. To the best of our knowledge, our work proposes the first active defense mechanism operatively to resist forking instead of following current recognitions passively.
  \item \textbf{Rigorous verifications.} Solid theoretical analyses and extensive simulations are conducted, which justify the validity of our forking models as well as the effectiveness of our active defense mechanism.
\end{itemize}

The rest of the paper is organized as follows. We first review the related work in Section \ref{sec:related work} and characterize the network topology of blockchain with defined expansivity of links in Section \ref{sec:system model}. Based on this, the activation time and activation degree are proposed in Sections \ref{section:activation time} and \ref{section:activation degree}. The modelings and analyses on the unintentional forking probability and intentional forking robust level are carried out in Sections \ref{section:model_un} and \ref{section:model_inten}. Our experimental results are shown in Section \ref{sec:exp}, and Section \ref{sec:conclusion} finally concludes our paper.

\section{Related work}\label{sec:related work}
At present, the growing interest in analyzing blockchain forking has produced a number of prominent investigations, which can be classified into two categories, i.e., the experiment-based and the model-based.

For the experiment-based forking analyses, Chen {\it et al.} \cite{chen} proposed a mechanism called GVScheme, which introduces the role of guarantor to effectively reduce the block verification time. And the simulation results manifest the efficacy of its mechanism in reducing the forking rate. Besides, a proximity-aware extension named Bitcoin Clustering Based Ping Time protocol (BCBPT) was presented by Fadihil {\it et al.} in \cite{proximity}, which alleviates the forking phenomenon via improving the transactions propagation process. BCBPT was designed to realize this through evaluating the ping latencies between nodes, which has been testified to be strongly effective in optimizing the performance through extensive simulations. In \cite{zhang}, Zhang {\it et al.} offered a backward-compatible defense mechanism to deal with selfish mining attack, allowing forking to be solved as a side problem. Through evaluating the proposed methods extensively, they concluded that the honest miners need to adopt the longest chain if it is at least 3 blocks ahead of
the competitors so as to fight back attackers.

Parallel to the above efforts, the model-based analysis heaved in sight by Decker {\it et al.} in \cite{P2P}, which quantitatively investigated the impact of information propagation on Bitcoin security from the macro level. To that aim, they proposed a probabilistic model to estimate the average forking rate based on the measurements of block spread across the network, and concluded that propagation delay causes forks primarily. Besides, Xiao {\it et al.} in \cite{P2P} presented an analytical model to evaluate the effect of network connectivity on consensus security where various adversaries are involved. As such, both the forking rate and the revenue of miners are established to elicit the security properties of consensus. To take precautions against forking, Wang {\it et al.} \cite{corking} studied the vulnerability of blockchain incurred by intentional forks through a delicate mathematical model, taking advantage of the large deviation theory. In their model, the number of created blocks from the honest nodes and adversaries are regarded as dynamically growing queues, in which the difference between them can be analyzed to reflect the vulnerability of blockchain. Different from current efforts, Chen {\it et al.} in \cite{impact} explored the impact of competition among mining pools on forking from the miners' perspective, based on which, a detailed model along with the derived closed-form formula of forking probability  and expected mining revenue of each pool was expressed. Moreover, an evolutionary game framework was then developed to further reveal the long-term trends in the computing power distributions over pools.

However, the above two kinds of studies fall short in the following detriments. On the one hand, the experiment-based analyses are limited to specific scenarios, making the conclusions drawn from a certain case may be unreliable in another. As a result, they can not support forking prevention from the top level since universal and decisive forking laws are inaccessible. On the other hand, existing model-based researches suffer from {\it spatiality-free} which means they are restricted to the representation of time dimension while ignoring the spatial characteristics of blockchain topology. However, forking is highly related to spatiality in the sense that the propagation of the main chain is carried out under a certain network topology. Therefore, the exact position and transmission capability of each node may exert a huge influence on forking. Additionally, the ongoing research only recognizes forking passively without proposing any active defense mechanism. Hence in this paper, we fill the gap through conducting forking analysis from the spatial-temporal dimension, based on which, a defense mechanism can be achieved proactively and operatively.
\section{System model}\label{sec:system model}
\subsection{Network topology}
\begin{figure}
  \centering
  \includegraphics[width=0.5\linewidth]{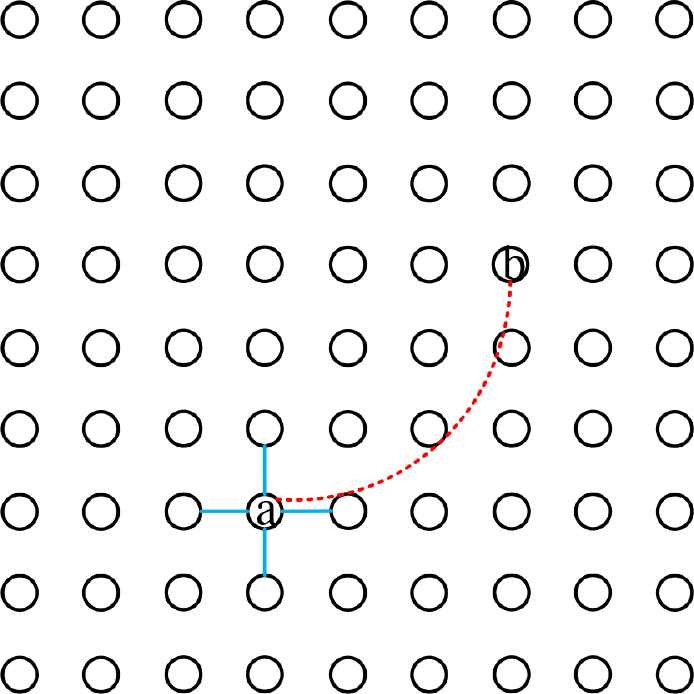}\\
  \caption{The topology of blockchain as a $9\times 9$ lattice for instance. To illustrate, nodes a and b are represented as $(4,3)$ and $(7,6)$ and the distance between them is 6. Note that the blue links are short-range links since the lattice distances between a and its neighbors are 1, and the red link is the long-range one since $d($a,$b)=6>1$. }\label{lattice}
\end{figure}

We exploit the two-dimensional grid network \cite{small} to abstractly establish the underlying network of blockchain, as a response to the pioneer theoretical developments which characterized the blockchain as a {\it small-world} phenomenon \cite{tnse,tao,ma}. Specifically, we deem the network as a graph $\mathbb{G}=(\mathbb{V},\mathbb{E})$, where $\mathbb{V}$ and $\mathbb{E}$ respectively denote the sets of nodes and edges. In particular, we signify the nodes engaging in blockchain as a set of lattice points, forming a $n\times n$ square as shown in Fig. \ref{lattice}. Each of them is identified by its location $\{(x,y):x\in\{1,...,n\},y\in\{1,...,n\}\}$. In doing so, the {\it lattice distance} between node $(x_1,y_1)$ and $(x_2,y_2)$ is determined to the number of ``lattice steps" detaching them: $d((x_1,y_1),(x_2,y_2))=|x_2-x_1|+|y_2-y_1|$. Notably, this grid model can be interpreted simply from the ``geographic" perspective, where nodes reside in the grid with neighbors in four directions. Hence, the location  represents the geographical position of a node and the lattice distance matches the spatial distance accordingly. Subsequently, we distinguish the links between any two nodes as the {\it short-range link} and {\it long-range link}, where the former specifies the link whose lattice distance of the two ends is 1 while the latter represents that of bigger than 1.

We claim that each node has four short-range links with neighbors determinately but has long-range links with some nodes probabilistically. In practice, the longer the spatial distance between two nodes, the less probability they have to possess a long-range link. Hence, we describe the probability that nodes $v_1$ and $v_2$ existing a long-range link, i.e., $P_{long}(v_1,v_2)$, retains a negative exponential relationship with distance. That is, $P_{long}(v_1,v_2)=d(v_1,v_2)^{-\beta}$ where $\beta$ is the long-range factor to ensure $d(v_1,v_2)^{-\beta}\in[0,1]$. Remarkably, such a designed factor envisions a quantifiable mechanism to defense forking actively, which will be illustrated detailedly in Sections \ref{section:model_un} and \ref{section:model_inten}.

\subsection{Layer propagation}
Once mining a new block, the creator will propagate it to all linked nodes, which then verify its validity and further transmit it to other connected nodes. Such a process continues until all the nodes accept this block, orchestrating the main chain propagation just like the {\it ripple effect}. Suppose the creator is node $s$, we then introduce the concept of {\it layer} to capture the spread-out characteristics of information propagation in blockchain by presenting the following definition.
\begin{definition}[Layer]
\begin{itemize}
  {\it \item Layer \#$0$: The set of nodes $\mathbb{V}_0$ belongs to layer \#$0$ if the lattice distance between node in $\mathbb{V}_0$ with $s$ equals 1.
  \item Layer \#$j,j\in N^+$: The set of nodes $\mathbb{V}_j$ belongs to layer \#$j$ if the lattice distance between node in $\mathbb{V}_j$ with $s$ is $(2^{j-1},2^j]$.}
\end{itemize}
\end{definition}

\begin{figure}
  \centering
  \includegraphics[width=0.5\linewidth]{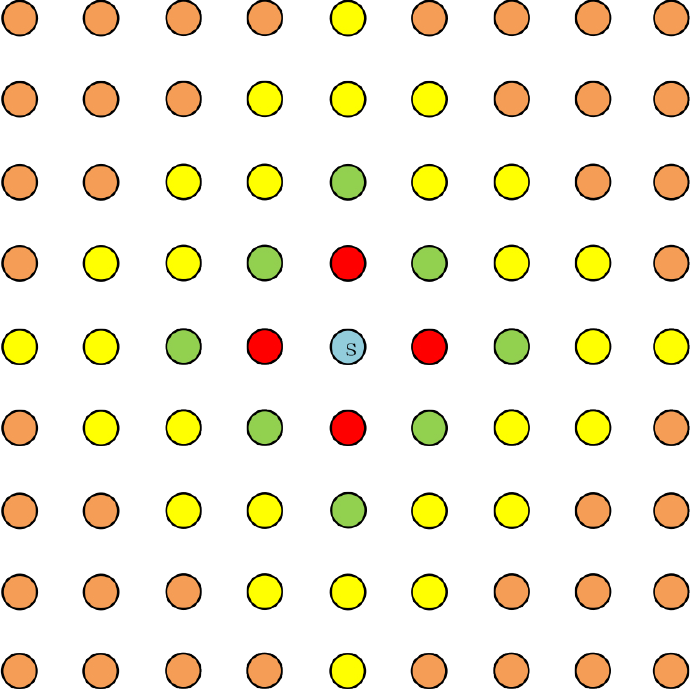}\\
  \caption{The illustration of layer. For example, layers \#1,2,3,4 contain the red, green, yellow, orange nodes, respectively. }\label{layer}
\end{figure}

Assume the longest lattice distance between any node with $s$ is $n_{max}\ge 2$, then we have $1\le j\le \log n_{max},j\in\mathbb{N}^+$. As shown in Fig. \ref{layer}, nodes with lattice distance $(1,2]$ from node $s$ compose layer \#1 (i.e., the green points), nodes with distance $(2,4]$ belong to layer \#2 (i.e., the yellow points), the same applies to other layers. If the farthest node receiving current main chain is $u$, and the lattice distance between $s$ and $u$ meets $2^{j-1}<d(s,u)\le 2^j$, then we state that $u \in \mathbb{V}_j$ and the main chain has been transmitted to layer \#$j$.

Recall the long-range link\footnote{For simplicity, we assume the long-range links only reside in adjacent layers, which means nodes in layer \#$j$ can only have long-range links with nodes in layer \#$j-1$ or \#$j+1$. Other long-range links between layer \#$j$ and \#$j-2/j+2$ or longer can be easily extended through our following descriptions.} probability is defined as $P_{long}(u,v)=d(u,v)^{-\beta}$ with the long-range factor $\beta$. One may conclude that a smaller $\beta$ induces a higher probability that the node can connect further, which in turn entitles a stronger potential to stretch more peripherally. That is to say, the long-range links also mirror various long-distance transmission capabilities, which brings in the following definition of {\it expansivity}.
\begin{definition}[Expansivity]\label{expansivity}  {\it The expansivity $\epsilon_L$ of  link $L\in\mathbb{E}$ with ends $u,v\in\mathbb{V}$ is defined as $\epsilon_L=P_{long}(u,v)\times \rho$  where $\rho$ denotes the  index of the layer where $v$ locates if $u$ is regarded as the creator.}
\end{definition}

According to Definition \ref{expansivity}, the expansivity of a short-range link equals 0 since $v$ belongs to layer \#$0$ in this case. As for a long-range link, its expansivity varies on the difference between $\beta$ and the lattice distance. Notably, the difference in expansivity indeed reflects connection heterogeneity.

Now we are in the position to highlight the advantages of the short-/long- range links to manifest the merits of our system model. To begin with, the short-range links exist between neighbors certainly, thus they offer great feasibility for {\it intra-layer} main chain propagation. On the contrary, the long-range link can probabilistically connect nodes when the lattice distance is larger than 1, which is often the case for nodes in different layers. Hence, the long-range links stand for the main pillar for {\it inter-layer} main chain propagation. To conclude, the short-/long- range links can cover various transmission cases of blockchain comprehensively. Additionally, the short-/long- range links with various expansivities are valuable for shaping the heterogeneity of links in blockchain. And it is the remote transfer ability of long-range links that breaks the regular pattern to propagate the main chain layer by layer, enabling cross-layer transmission. This facilitates a more pragmatic and realistic blockchain.


However, the combination of short-/long- range links challenges the rigorous inspection of main chain propagation. This is because the cross-layer transmission enabled by long-range links may accelerate information propagation with great uncertainty for inter-layer cases, perplexing the transmission dynamics. 
To remedy this, we first quantify the time when each layer \#$j$ is first noticed by the main chain, i.e., the activation time $\mathbb{E}[T(j)]$. Based on this, we can derive the activation degree $I(t)$, defined as the number of nodes for each layer that has perceived the main chain at time $t$, which finally realizes our main chain propagation model.




\section{Activation time}\label{section:activation time}
In this section, we model the {\it activation time} of each layer to describe the absolute time when the nodes in each layer first accept the main chain. To achieve this, we carry out a {\it relative-to-absolute} method, which starts by introducing the {\it adjacent propagation time} to denote the relative time difference of layer \#$j$ getting the main chain from layer \#$j-1$. After that, we deduce the corresponding absolute time by summing all the cases.

\begin{definition}[Adjacent propagation time]
{\it The adjacent propagation time (APT) $T_j$ refers to the arriving time that the main chain reaches layer \#$j$, if the time of layer \#$j-1$ receives the main chain is set as 0.}
\end{definition}

Accordingly, APT denotes the interval of main chain propagation from layer \#$j-1$ to \#$j$. Suppose the average transmission delays via short-/long- range links are denoted as $\Delta_s, \Delta_l$, where $\Delta_l\ge\Delta_s$ holds obviously. Then we can demonstrate APT for each layer in the following theorem.
\begin{theorem}\label{theo1}
{\it The expected adjacent propagation time of each layer \#$j\ge 0$, i.e., $\mathbb{E} [T_j]$, can be summarized as
\begin{equation}\label{APT1}
\left\{
             \begin{array}{lr}
            \mathbb{E} [T_j]=\Delta_s, & j=0,1 , \\
             \mathbb{E} [T_j]>\tau_l^j(1-(1-(3\times2^{j-1})^{-\beta})^{f_j})\\
  +c\Delta_s\cdot 2^{j-2}(1-(3\times2^{j-1})^{-\beta})^{f_j}, & j>1,
             \end{array}
\right.
\end{equation}
where $\tau_l^j$ denotes the time that layer \#$j$ gets the main chain from \#$j-1$ via the long-range links, and $f_j=N_{j-1}N_j$ with $N_{j-1}$ and $N_j$ being the number of nodes in  $\mathbb{V}_{j-1}$ and $\mathbb{V}_j$.}
\end{theorem}

\begin{IEEEproof} When $j=0$, the lattice distance between nodes in $\mathbb{V}_0$ to node $s$ is 1, hence the propagation interval from $s$ to $\mathbb{V}_0$ is $\Delta_s$ via the short-range link; when $j=1$, the lattice distance between nodes in $\mathbb{V}_1$ to node $s$ is 2, indicating that of between nodes in $\mathbb{V}_1$ to nodes in $\mathbb{V}_0$ is 1. Hence the propagation interval from $\mathbb{V}_0$ to $\mathbb{V}_1$ is $\Delta_s$ via direct short-range link due to $\Delta_s\le\Delta_l$. This justifies the first line of \eqref{APT1}.

Since the short-/long- range links exist between nodes when their lattice distance is 1 or in adjacent layers, the main chain can be transmitted to layer \#$j$ from \#$j-1$ via both of them. Hence, the APT of layer \#$j\ge1$ can be characterized as \begin{equation}\label{APT}
\mathbb{E}[T_j]=P_l^j \cdot \tau_l^j+P_s^j \cdot \tau_s^j,
\end{equation}
where $P_l^j$ and $\tau_l^j$ denote the probability and time that layer \#$j$ gets the main chain from \#$j-1$ via the long-range links, and $P_s^j$ and $\tau_s^j$ express those of via the short-range links. We name $P_l^j \cdot \tau_l^j$ and $P_s^j \cdot \tau_s^j$ respectively as the {\it long-range} and {\it short-range} parts which will be demonstrated more detailedly in the following.

(a) {\it The long-range part.} Denote the probability of nodes $u$ and $v$ residing in layer \#$j-1$ and \#$j$ having the long-range link as $p_{uv}$, then we have $p_{uv}=d(u,v)^{-\beta}>(3\times2^{j-1})^{-\beta}$ because of $d(u,v)<2^{j-1}+2^j=3\times2^{j-1}$. Hence, the probability that node $u$ does not possess long-range links with any node in layer \#$j$ can be specified as $\bar{p}_{u\rightarrow\#j}=\prod \limits_{v\in\mathbb{V}_j} (1-p_{uv})$. Let the number of nodes in $\mathbb{V}_j$ is $|\mathbb{V}_j|=N_j$, then we can obtain $N_j=\sum_{i=2^{j-1}+1}^{2^j} 4i=3\times2^{2j-1}+2^j$ since there are $4i$ nodes whose lattice distance from node $u$ is $i$. In light of this, we can get
\begin{equation}\label{pu-j}
\begin{aligned}
\bar{p}_{u\rightarrow\#j}&=\prod\limits_{v\in\mathbb{V}_j} (1-p_{uv})<\prod\limits_{n_v=1}^{N_j} (1-(3\times2^{j-1})^{-\beta})\\
&=(1-(3\times2^{j-1})^{-\beta})^{N_j}.
\end{aligned}
\end{equation}

Accordingly, the probability that any node in layer \#$j-1$ can not connect any node in layer \#$j$ with long-range link can be formalized as $\bar{p}_{\#j-1\rightarrow\#j}=\prod\limits_{u\in\mathbb{V}_{j-1}} \bar{p}_{u\rightarrow\#j}=\prod\limits_{u\in\mathbb{V}_{j-1}}\prod\limits_{v\in\mathbb{V}_j} (1-p_{uv})$. Based on $N_{j-1}=3\times2^{2j-3}+2^{j-1}$ and suppose $f_j=N_{j-1}N_j=9\times2^{4j-4}+9\times2^{3j-3}+2^{2j-1}$, we can conclude that
\begin{equation}\label{equ}
\begin{aligned}
\bar{p}_{\#j-1\rightarrow\#j}<(1-(3\times2^{j-1})^{-\beta})^{f_j}.
\end{aligned}
\end{equation}
Hence, the probability that any node in layer \#$j-1$ has a long-range link with any node in layer \#$j$ can be denoted as $p_{\#j-1\rightarrow\#j}=1-\bar{p}_{\#j-1\rightarrow\#j}>1-(1-(3\times2^{j-1})^{-\beta})^{f_j}$, which indicates that $P_l^j>1-(1-(3\times2^{j-1})^{-\beta})^{f_j}$.

Notably, layer \#$j$ receives the main chain from layer \#$j-1$ via the long-range link implies that during network transmission in layer \#$j-1$, there is a specific node (denoted as $u^*$) that has accepted the main chain, and also possesses a long-range link with nodes in layer \#$j$ exactly. As a result, node $u^*$ propagates the main chain to the next layer via this valuable long-range link. Let the first node accepting the main chain in layer \#$j-1$ be $\bar{u}$. If $\bar{u}$=$u^*$, then $\tau_l^j=\Delta_l$. Otherwise, $\tau_l^j=d(u^*,\bar{u})\Delta_s+\Delta_l$ since node $u^*$ needs to receive the main chain from node $\bar{u}$ through the short-range links internally. In this paper, we simply utilize the term $\tau_l^j$ for demonstration and leave an exact analysis of $\tau_l^j$ for future work. Until now, we can deduce the APT of layer \#$j$ via the long-range link as
\begin{equation}
\begin{aligned}
\mathbb{E}[T_j^l]=P_l^j \cdot \tau_l^j>\tau_l^j(1-(1-(3\times2^{j-1})^{-\beta})^{f_j}).
\end{aligned}
\end{equation}

(b) {\it The short-range part.}
Since the short-range links only possess in neighbors whose lattice distance is 1, when the nodes in layer \#$j-1$ propagate the main chain to the next layer through the short-range links, the distance to be passed will be no more than $2^{j-1}-2^{j-2}=2^{j-2}$. Hence, we can set $\mathbb{E}[\tau_s^j]=c\cdot \Delta_s 2^{j-2}$ with $c$ as the discounting factor to get the expectation. With this in mind, we can get the APT of layer \#$j$ via the short-range link as
\begin{equation}
\begin{aligned}
\mathbb{E}[T_j^s]=P_s^j \cdot \tau_s^j=c\cdot \Delta_s 2^{j-2}(1-P_l^j).
\end{aligned}
\end{equation}

In summary, we can get the APT of layer \#$j>1$ as shown in the second line of \eqref{APT1} through combining the above two parts. This completes our proof.
\end{IEEEproof}

It is worth noting that the defined APT only focuses on the propagation time of the main chain between nearby layers. To derive the whole expected propagation time of the main chain from the source node $s$ to layer \#$j$, i.e., $\mathbb{E}[T(j)]$, it is necessary to sum all the corresponding cases as presented in the following:
\begin{equation}\label{index}
\mathbb{E}[T(j)]=
\left\{
             \begin{array}{lr}
            \mathbb{E}[T_0], & j=0, \\
             \mathbb{E}[T_0]+\mathbb{E}[T_1], & j=1,\\
             \mathbb{E}[T_0]+\mathbb{E}[T_1]+\sum\limits_2^j\mathbb{E}[T_j], & j>1.
             \end{array}
\right.
\end{equation}
Through substituting \eqref{APT1} into \eqref{index}, the expected activation time when the nodes in each layer \#$j$ receive the main chain firstly can be summarized as
\begin{equation}\label{tj}
\mathbb{E}[T(j)]=
\left\{
             \begin{array}{lr}
            \Delta_s, & j=0, \\
             2\Delta_s, & j=1,\\
             2\Delta_s+\sum\limits_2^j\mathbb{E}[T_j], & j>1,
             \end{array}
\right.
\end{equation}
in which $\mathbb{E}[T(j)]>2\Delta_s+\sum\limits_2^j[\tau_l^j(1-(1-(3\times2^{j-1})^{-\beta})^{f_j})+c\Delta_s\cdot 2^{j-2}(1-(3\times2^{j-1})^{-\beta})^{f_j}]$.
\section{Activation Degree}\label{section:activation degree}
This section will illustrate the main chain propagation model by specifying the number of nodes that have received the main chain at any time, which is also denoted as the {\it activation degree}, based on the above analyses on {\it activation time}. To this end, we take advantage of the epidemic model, i.e., the susceptible-infected (SI) model \cite{SI}, to characterize the evolution of nodes in blockchain. SI models are widely exerted in topologically related information diffusion phenomenons, both in social and technology networks \cite{SI1,SI2, SI3, corking}. We believe it is the simplest method to capture the essence of information propagation in blockchain. Technically, nodes have two states: susceptible ($\mathbb{S}$) and infected ($\mathbb{I}$), which are interchangeable. A node in state $\mathbb{S}$ indicates it is inactivated that has not received the main chain while that in state $\mathbb{I}$ means it has been activated by accepting the main chain. A node in $\mathbb{S}$ may turn into $\mathbb{I}$ once it accepts the main chain and work on it. In the following, we first derive the activation degree within layer locally, after which, we can obtain the global degree in the network scale.

We set the beginning of main chain diffusion in layer \#$j>1$ as time $T=0$, and the activation and inactivation degrees (or the number of nodes that are in state $\mathbb{I}$ and $\mathbb{S}$) are denoted as $I_j(\omega)$ and $S_j(\omega)$ respectively in any time $T=\omega>0$. Then we have $I_j(\omega)+S_j(\omega)=N_j$ and $I_j(0)=1$ since there must be one activated node initially. Based on the SI model, $I_j(\omega)$ can be derived as:
\begin{equation}\label{I(t)}
\begin{aligned}
\left\{
             \begin{array}{lr}
            \frac{dI_j(\omega)}{d\omega}=\alpha_j I_j(\omega)(N_j-I_j(\omega)),\\
             I_j(0)=1,
             \end{array}
\right.
\end{aligned}
\end{equation}
where $\alpha_j$ represents the probability that an activated node transfers the main chain to others within layer \#$j$. Considering nodes inside the layer transfer information via short-range links, $\alpha_j$ can also be interpreted as the probability that a node possesses the short-range link with others, which is $\alpha_j=\frac{4}{N_j-1}$. In doing so, we can derive $I_j(\omega)$ formally through the logistic model, leading to
\begin{equation}\label{I(t)2}
\begin{aligned}
I_j(\omega)=\frac{N_j}{1+(N_j-1)\mathrm{e}^{-\alpha_jN_j\omega}}.
\end{aligned}
\end{equation}

Notably, $I_j(\omega)$ in \eqref{I(t)2} depicts the number of activated nodes in time $\omega$ of layer \#$j$ locally. Subsequently, we are going to find the activation degree globally, i.e., $I(t)$, which expresses the number of infected nodes in the network at $T=t$ from the very beginning when the source node produces the main chain. We reset $T=0$ as the time when the main chain is originally generated, in light of the above analysis, we can acquire the following theorem.
\begin{theorem}
{\it The activation degree of the network
$I(t)$ when $0<t<T(J),J\in[0,\log n_{max}],J\in\mathbb{N}^+$, satisfies}
\begin{equation}\label{I(tt)}
\begin{aligned}
I(t)=
\left\{
             \begin{array}{lr}
            1, & 0<t<T(0), \\
             5, & T(0)\le t<T(1),\\
             13, & T(1)\le t<T(2),\\
             13+\sum\limits_{j=3}^{J-1}I_j(t-T(j)), & T(2)\le t<T(J).
             \end{array}
\right.
\end{aligned}
\end{equation}
\end{theorem}

\begin{IEEEproof}
We prove each case of \eqref{I(tt)} in the following.

(1) Case 1: The activation degree of the network
$I(t_0)=1$ since layer \#$0$ has not received the main chain when $t_0\in(0,T(0))$ according to \eqref{index}, shown as the blue node in Fig. \ref{layer}.

(2) Case 2: The activation degree of the network
$I(t_1)=5$ since layer \#$0$ can receive the main chain when $t_1\in[T(0),T(1))$, leading to the four directly connect neighbors being activated, shown as the blue and red nodes in Fig. \ref{layer}.

(3) Case 3: The activation degree of the network
$I(t_2)=13$ since both layer \#$0$ and \#$1$ can receive the main chain when $t_2\in[T(1),T(2))$, making $I(t_2)=I(t_1)+N_1=13$, shown as the blue, red and green nodes in Fig. \ref{layer}.

(4) Case 4: when $t_j\in[T(2),T(J)), 2<j\le J$, layers \#$0\sim$\#$j-1$ have all accepted the main chain. Take layer \#$3$ as an example, the activation time of this layer is $T(3)$. Hence, the main chain has been propagated within this layer for $\triangle t_3=t_j-T(3)$, making the activation degree in this layer as $I_3(\triangle t_3)$. Other layers can be done in the same manner, leading to the fact that $I(t_j)=I(t_2)+\sum_{j=3}^{J-1}I_j(\triangle t_j)$.
\end{IEEEproof}


Until now, we present the activation degree at any time in the network, which can be recognized as the main chain propagation model. Such a measurement is so valuable in reflecting the possible forking occurrence since the higher the degree is, the more nodes perceive the main chain, the less probability that forking may happen. In the following two sections, we carry out forking analysis for both unintentional and intentional scenarios, based on which, an active defense mechanism can be proposed.

\section{Modeling and Analysis: Unintentional forking}\label{section:model_un}
Technically, unintentional forking is normally caused by the inefficient propagation of the blockchain overlay network without any attacker or artificial manipulation \cite{black},\cite{liu}. Trace the root, it happens when a conflict block $B$ is found while the previous block $A$ with the same height is propagated in the network. Accordingly, we claim that the unintentional forking probability is equivalent to that of generating new blocks (i.e., $B$) during the propagation of chain containing $A$.

Assume the source node starts to transmit the produced main chain at time  $t=0$, and all the nodes are supposed to accept the main chain at $t=T_{all}$ if no forking occurs. The whole time span of $T_{all}$ can be equally divided into $\frac{T_{all}}{\delta}$ intervals, with each interval being $\delta$. In the following, we proceed to analyze the probability of unintentional forking both in unit time, i.e., $fr(t)$, and over a period of time, i.e., $FR(t)$, where the former focuses on each interval $[t,t+\delta]$ while the latter specifies any defined period $[0,t+\delta]$
with $t\in[0,T_{all}-\delta]$. In doing so, effective countermeasures of unintentional forking are facilitated operatively at different time scales with multiple intensities.
\subsection{Unintentional forking probability analysis in unit time}\label{a}
In fact, the mainstream consensus mechanism, i.e., {\it proof-of-work} (PoW), guarantees that the average work of each node follows Bernoulli distribution, which suggests that the mining process can be deemed as a Poisson process \cite{corking, black}. Let $\Omega(t)$ be the set of nodes which have not received the main chain at time $t$ and $\varphi(t)$ denotes the number of conflict blocks created by the nodes in $\Omega(t)$ at time $t$. Hence, $\varphi(t)$ follows Poisson distribution with expectation $\lambda=\sum_{u\in\Omega(t)}\pi_u$, with $\pi_u$ denoting the computing power of node $u$. As such, we can obtain the probability that $k$ new blocks are generated during $[t,t+\delta]$ as
\begin{equation}
\begin{aligned}
P[\varphi(t+\delta)-\varphi(t)=k]=\frac{\mathrm{e}^{-\lambda\delta}(\lambda\delta)^k}{k!}.
\end{aligned}
\end{equation}

Note that $k=0$ elaborates there is no block created by the inactivated nodes in $\Omega(t)$, which in turn indicates no unintentional forking occurrence. Therefore, we can derive the probability of unintentional forking as $fr(t)=1-P[\varphi(t+\delta)-\varphi(t)=0]
=1-\mathrm{e}^{-\delta\sum_{u\in\Omega(t)}\pi_u}$. If the computing power is uniformly distributed among nodes, we have
\begin{equation}\label{fr(t)}
\begin{aligned}
fr(t)=1-\mathrm{e}^{-\delta|\Omega(t)|\pi_u}=1-\mathrm{e}^{-\delta S(t)\pi_u},
\end{aligned}
\end{equation}
where $S(t)=N_{total}-I(t)$ with $N_{total}=\sum_{i=1}^n 4i+1=1+2n+n^2$ and $I(t)$ is formalized in \eqref{I(tt)}. Through investigating \eqref{fr(t)} thoroughly, we claim the following corollaries to reveal the inherent mechanism of unintentional forking in unit time, thus we can present direct evidence to hinder it operatively.
\begin{corollary}\label{c1}
{\it A smaller long-range factor $\beta$ can induce a lower $fr(t)$.}
\end{corollary}
\begin{IEEEproof}
As described above, $\beta$ specifies the probability that any node pair $u$ and $v$ has a long-range link. The lower $\beta$ is, the more likely such link will exist. In addition, the long-range link facilitates the synchronicity of cross-layer main chain propagation, which allows more nodes to receive the main chain earlier, making the number of suspectable nodes smaller. Aware of this, we can clarify that a lower $\beta$ enables the main chain to be broadcasted to the whole network more efficiently through more long-range links, shrinking the unintentional forking probability as a consequence.
\end{IEEEproof}

\begin{corollary}\label{c2}
{\it The unintentional forking probability $fr(t)$ goes down when the transmission delays of the short-/long- range links, i.e., $\Delta_s$ and $\Delta_l$, decline.}
\end{corollary}
\begin{IEEEproof}
As demonstrated in Theorem \ref{theo1} and \eqref{tj}, a lower $\Delta_s/\Delta_l$ will decrease the expected APT as well as the activation time for each layer \#$j$. This will in turn influence $I(t)$ in \eqref{I(tt)} positively while  affect $S(t)$ negatively, indicating a lower forking probability as formalized in \eqref{fr(t)}. An intuition for this is that lower transmission delays make blockchain consume less time to hear the main chain for the inactivated nodes. This leaves no room for them to produce competitive blocks at the same height, decreasing forking probability consequently.
\end{IEEEproof}

\subsection{Unintentional forking probability analysis over a period of time}
In this section, we formalize the probability of unintentional forking during a period of time $[0,t+\delta]$, i.e., $FR(t)$. According to subsection \ref{a}, the forking probability in $[t,t+\delta]$ is denoted as $fr(t)$, then $FR(t)$ turns to
\begin{equation}\label{FR(t)}
\begin{aligned}
FR(t)=1-\prod_{w=0}^t(1-fr(w))=1-\prod_{w=0}^t\mathrm{e}^{-\delta\sum_{u\in\Omega(w)}\pi_u}.
\end{aligned}
\end{equation}
When the computing power is allocated uniformly, $FR(t)$ becomes $1-\mathrm{e}^{-\delta\pi_u\sum_{w=0}^t S(w)}$. We then give the following corollaries.
\begin{corollary}\label{c3}
{\it A smaller long-range factor $\beta$ can bring about a lower $FR(t)$.}
\end{corollary}
\begin{corollary}\label{c4}
{\it $FR(t)$ drops when $\Delta_s$ and $\Delta_l$ decrease.}
\end{corollary}

The proofs of these two corollaries are similar to those of Corollaries \ref{c1} and \ref{c2}, so we omit them for simplicity.
\section{Modeling and Analysis: Intentional forking}\label{section:model_inten}
Differently with unintentional forking, intentional forking involves malicious adversaries, who aim to surreptitiously execute the PoW for creating a chain containing fake transactions, namely the {\it fake chain} \cite{corking}. If the fake chain is longer than the honest main chain\footnote{In this section, we differentiate the chains created by the honest nodes and the adversaries as the honest main chain and the fake chain, respectively.}, the latter is then substituted by the malicious fake one according to the longest chain principle, incurring intentional forking attack as a consequence.

In particular, the adversaries always commence the forking attack assembly as a group since the honest majority makes it challenging for an attacker to commit attacks solely. In such an adversarial group, members share knowledge of the fake chain and information interchange is required to be real-time so as to prompt the success probability of attack.  Considering this, we can assume there is no transmission delay among the group which can be realized through maintaining the participants in a local area network (LAN). Based on this, Wang {\it et al.} in \cite{corking} presented the robust level $\vartheta$ of blockchain being attacked by intentional forking through analyzing the computational confrontation between the honest and malicious nodes, that is
\begin{equation}
\begin{aligned}
\vartheta=-\frac{\log \epsilon}{\log \frac{\gamma m}{\Lambda e}},
\end{aligned}
\end{equation}
where $\epsilon\in[0,1]$ is a preset threshold indicating the vulnerability probability of blockchain should be less than. In addition, $m$ and $e$ describe the number of nodes that have received the honest main chain and fake chain, $\gamma$ and $\Lambda$ represent the computing power of the honest and malicious node on the premise of uniform distribution. Essentially, $\vartheta$ suggests the required difference in the number of blocks completed by the adversary and the honest node, to verify a transaction. The higher $\vartheta$ is, the more effort should be put to launch attacks, the safer the blockchain being protected from intentional forking.

However, the above formulation falls short in {\it static modeling} and {\it spatiality-free}, making it insufficient in depicting a real-time and pragmatic blockchain. Aware of this, we present a dynamic robust level with spatiality, i.e., $\theta(t)$ as follows:
\begin{equation}
\begin{aligned}
\theta(t)=-\frac{\log \epsilon}{\log \frac{\gamma I(t)}{\Lambda e}},
\end{aligned}
\end{equation}
where $I(t)$ is determined as \eqref{I(tt)}. This gives the following corollaries.
\begin{corollary}\label{c5}
{\it A smaller long-range factor $\beta$ can lead to a higher $\theta(t)$, thus making blockchain stronger in fighting against intentional forking.}
\end{corollary}
\begin{corollary}\label{c6}
{\it $\theta(t)$ raises when $\Delta_s$ and $\Delta_l$ decrease, which indicates that dwindling the transmission delays can help to hinder intentional forking.}
\end{corollary}

These two corollaries can be easily derived, therefore we omit their proofs for simplicity.

\textbf{Remark}. Here we are going to highlight the significant merits of all the proposed corollaries in facilitating an active defense mechanism for thwarting forking. Note that for both unintentional and intentional scenarios, setting a smaller $\beta$ and $\Delta_s/\Delta_l$ will lead to a lower forking probability and promote the robustness of blockchain, which means 1) increasing the long-range link between any two nodes and 2) dwindling the transmission delays of the short-/long- range links are the two effective countermeasures. This provides clear evidence for blockchain designers to devise and detect the topology and transmission capacity of the network so as to meet our requirements in fighting back forking. More excitingly, our paper proposes the first {\it active} forking defense mechanism, that is, {\it positively reshaping the overlay network through decreasing the long-range factor $\beta$.} In doing so, we can eradicate the forking phenomenon fundamentally.


\section{Experiments}\label{sec:exp}
In this section, we will evaluate the validity and effectiveness of our proposed models and active defense mechanism. We have simulated the blockchain network with the PoW consensus mechanism written by {\it Python}, running on machines equipped with Intel Core i7-8700 GPU, 3.20 GHz CPU and 8 GB RAM. We begin by verifying our main chain propagation model in subsection \ref{ex-A}, after that, the unintentional and intentional forking models established in Sections \ref{section:model_un} and \ref{section:model_inten} are analyzed respectively in subsections \ref{ex-B} and \ref{ex-C}.
\subsection{Evaluation on the main chain propagation model}\label{ex-A}
We first carry out experiments on the activation degree $I(t)$ and compare it with the theoretical results, to demonstrate the validity of our proposed models. To calculate $I(t)$, we simulate the topology of blockchain as the aforementioned grid of $N_{total}=145$ and 545 nodes with short-/long- range links. Besides, the probability of the long-range link between any node pair $u$ and $v$ is preset as $P_{long}=d(u,v)^{-\beta}$ where $\beta=1$ and 10. The transmission delays are determined as $\Delta_s=1$ and $\Delta_l=1.5$ respectively. Note the tested system is driven by PoW consensus and we fix the interval $\delta$ as $\Delta_s$ for better performance. Each experiment is repeated 50 times to get the average result for statistic confidence.

\begin{figure}
	\centering
	\subfigure[$\beta=1$, $N_{total}=145$]{
		\begin{minipage}[t]{0.45\linewidth}
			\centering
			\includegraphics[width=1.68in]{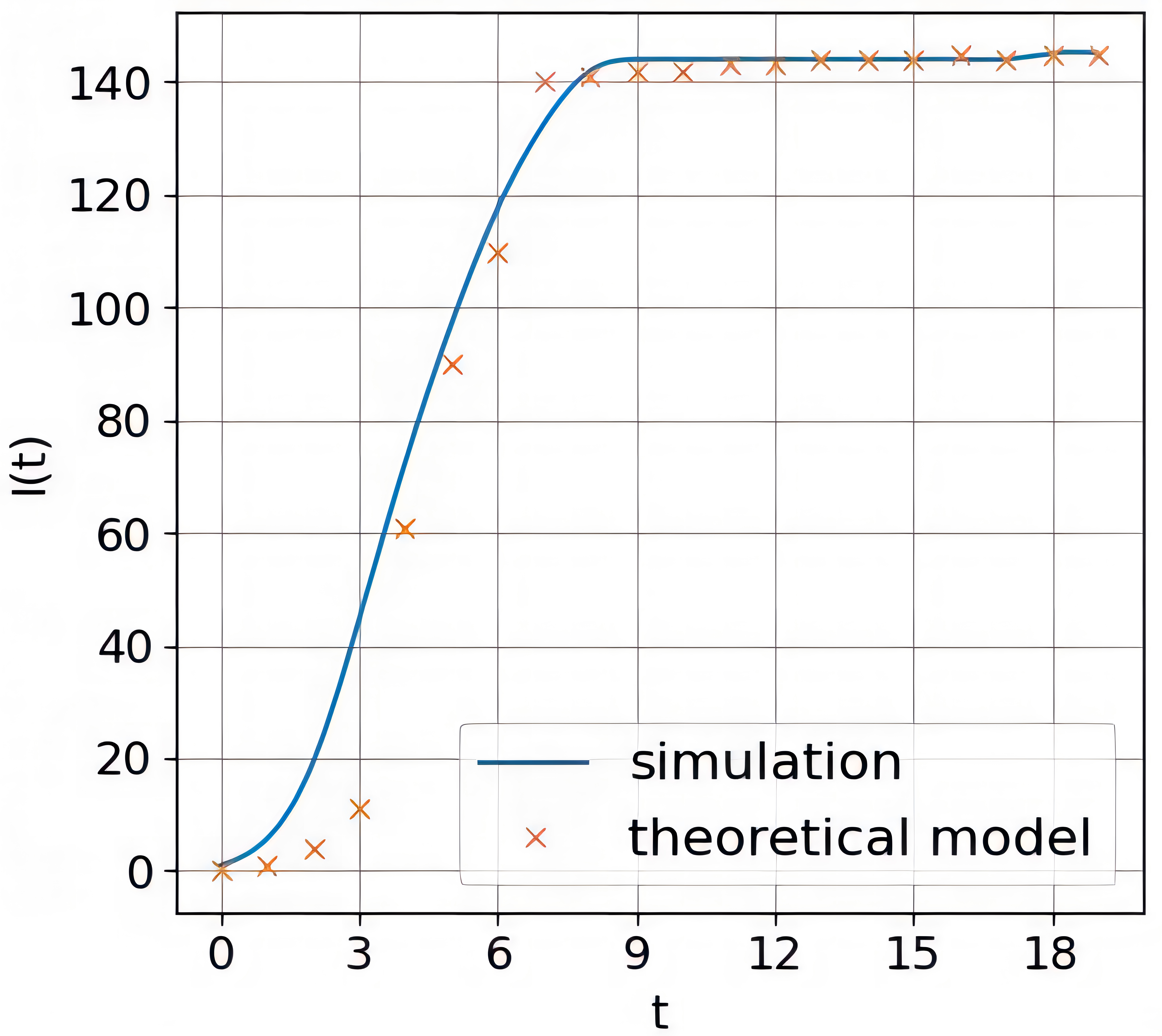}
		\end{minipage}
	}%
	\subfigure[$\beta=1$, $N_{total}=545$]{
		\begin{minipage}[t]{0.45\linewidth}
			\centering
			\includegraphics[width=1.68in]{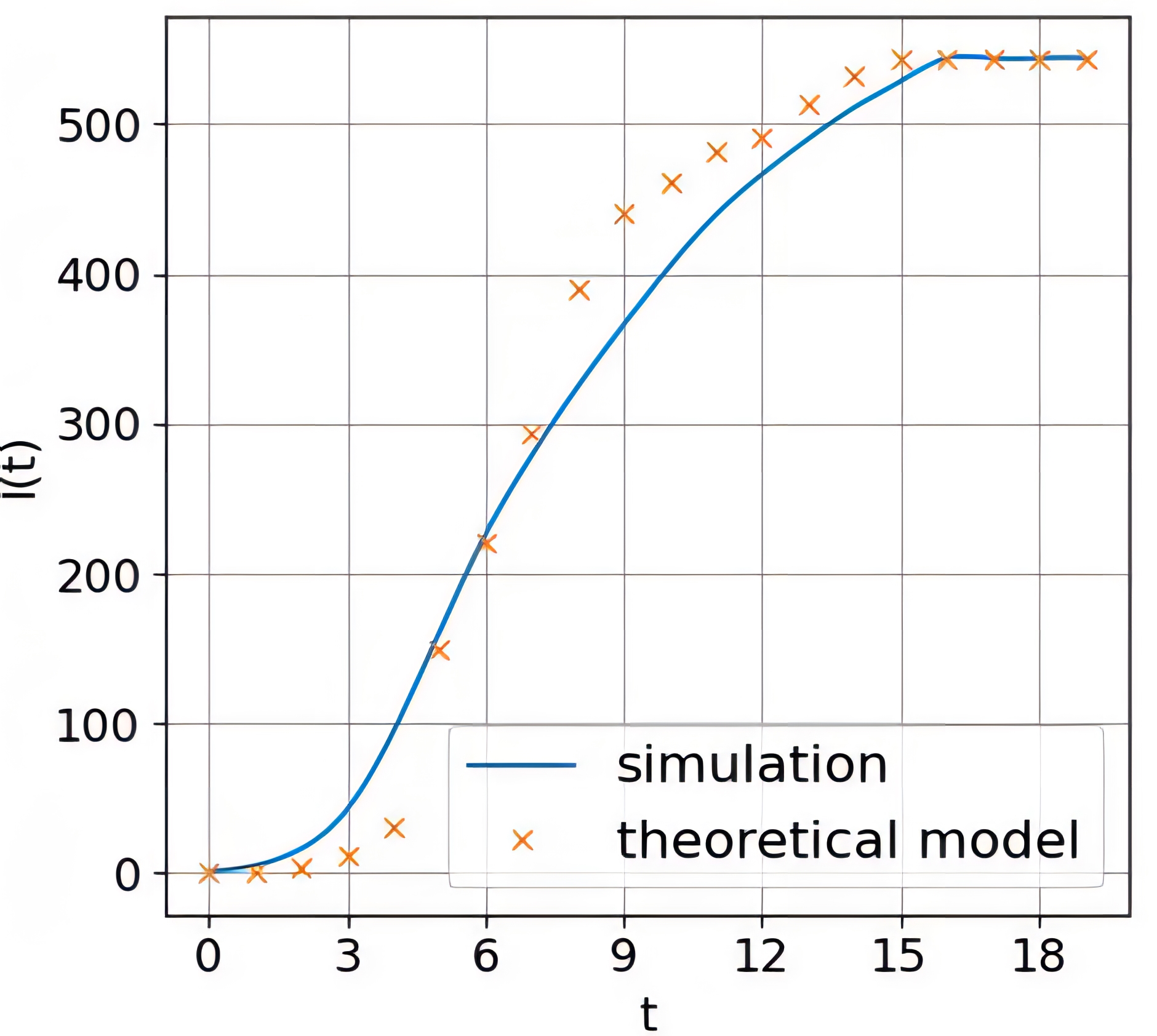}
		\end{minipage}
	}%

	\subfigure[$\beta=10$, $N_{total}=145$]{
		\begin{minipage}[t]{0.45\linewidth}
			\centering
			\includegraphics[width=1.68in]{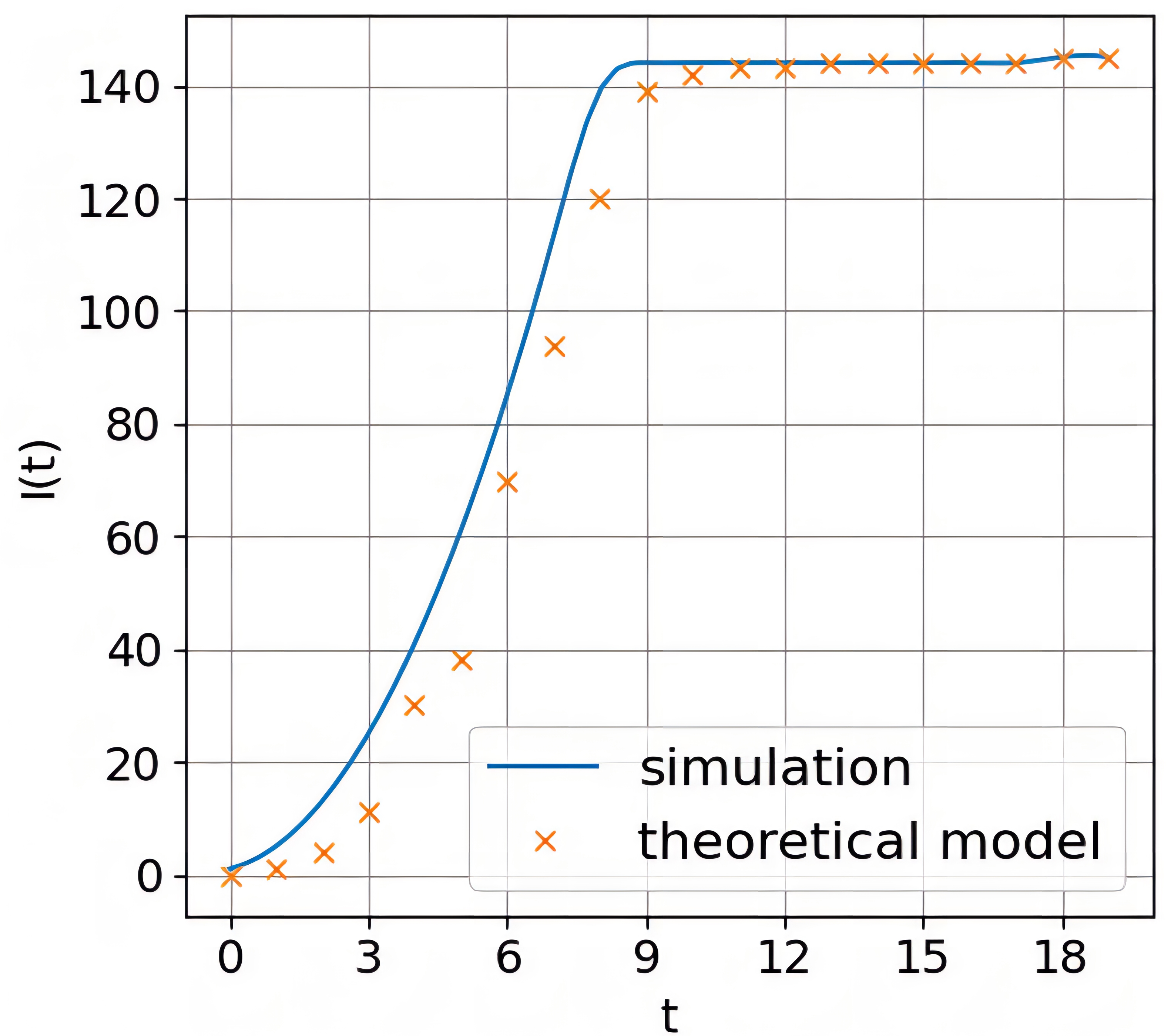}
		\end{minipage}
	}%
	\subfigure[$\beta=10$, $N_{total}=545$]{
		\begin{minipage}[t]{0.45\linewidth}
			\centering
			\includegraphics[width=1.68in]{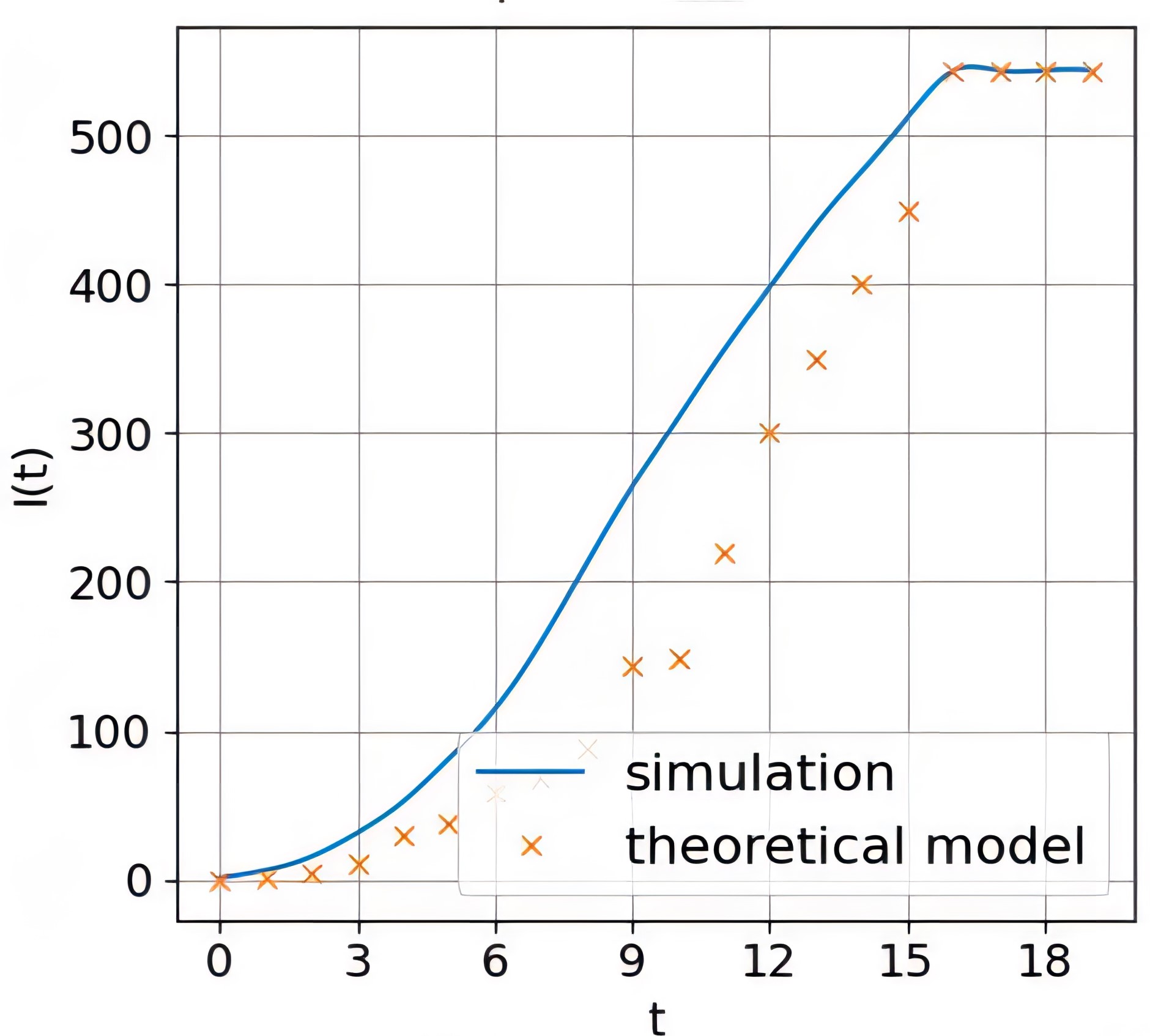}
		\end{minipage}
	}%
	\centering
	\caption{The comparison between experimental and analytical results on the activation degree $I(t)$.}
	\label{fig:I(t)}
\end{figure}

Fig. \ref{fig:I(t)} reports the theoretical and simulated results of $I(t)$ on the difference of $t$  when $\beta=1$ and 10 and $N_{total}=145$ and $545$. Conclusively, we make the following observations: 1) our experimental and analytical results match perfectly under various network sizes and topologies. This demonstrates the validity of our main chain propagation model in depicting the information diffusion of blockchain. 2) From subfigures (a) and (c), we can conclude that under the same network size with $N_{total}=145$, the increase of $\beta$ will impose a negative effect on $I(t)$. More specifically, when $t=6$, the activation degree $I(t)$ of $\beta=1$ is nearly 120 in the simulation (or 110 theoretically), while that of $\beta=10$ is 80 approximately in the simulation (or 70 theoretically). One conjecture is that a higher $\beta$ may trigger a lower probability of possessing a long-range link between any nodes $u$ and $v$, slowing down the propagation process cross-layer and consequently declining the number of activated nodes. The same conclusion can also be acquired from subfigures (b) and (d). 3) From subfigures (a) and (b), we can find that when $\beta$ is identical, the network with $N_{total}=145$ realizes consensus globally earlier than that of with $N_{total}=545$, which reports a smaller network is much easier to reach consistency. Detailedly, all the nodes get the main chain when $t=9$ in the network with 145 nodes, while $I(t)$ in the network with 545 nodes stops raising until $t=17$. The same can be concluded from subfigures (c) and (d).

\begin{figure}
	\centering
	\subfigure[$N_{total}=145$]{
		\begin{minipage}[t]{0.45\linewidth}
			\centering
			\includegraphics[width=1.65in]{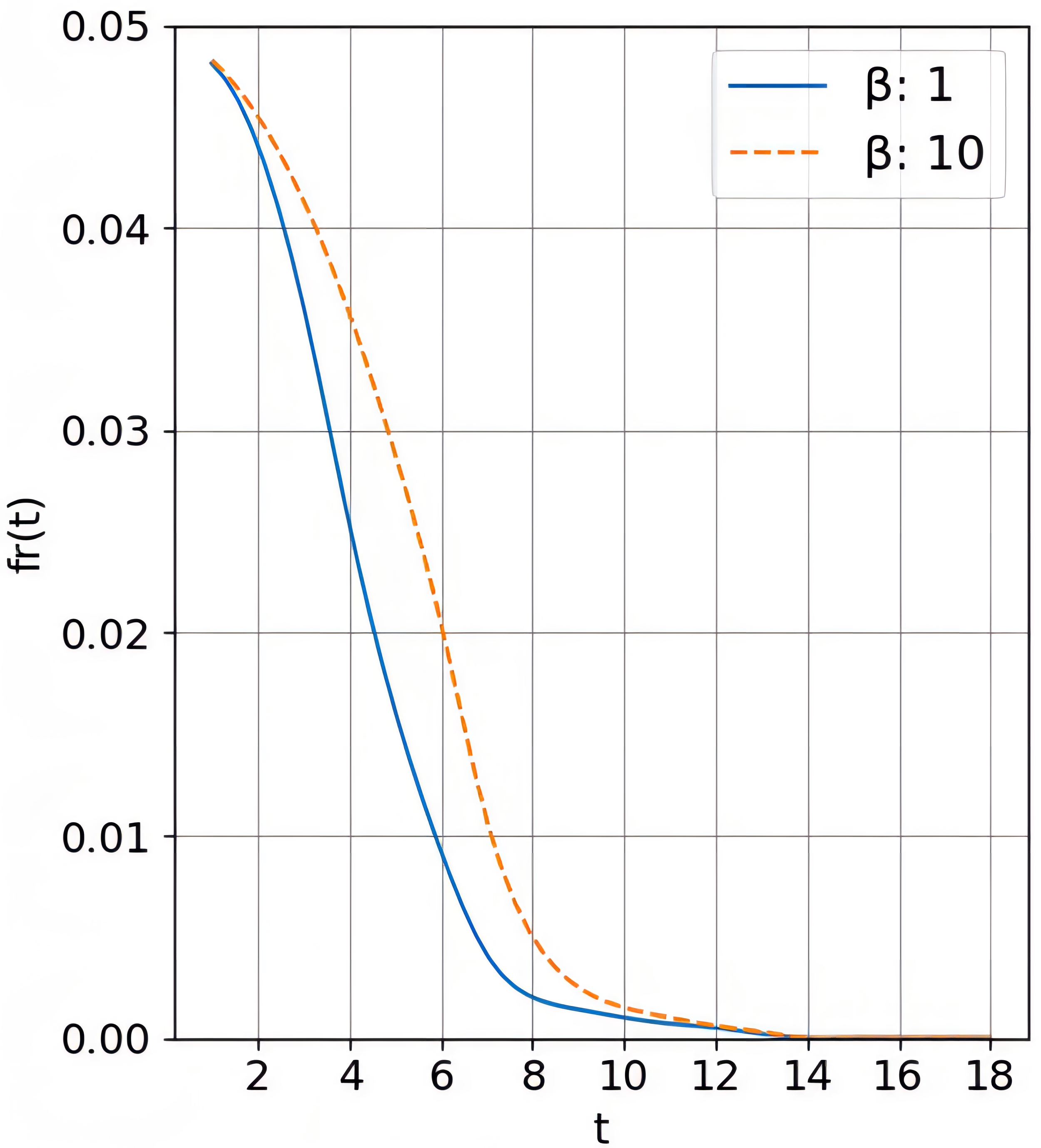}
		\end{minipage}
	}%
	\subfigure[$N_{total}=545$]{
		\begin{minipage}[t]{0.45\linewidth}
			\centering
			\includegraphics[width=1.68in]{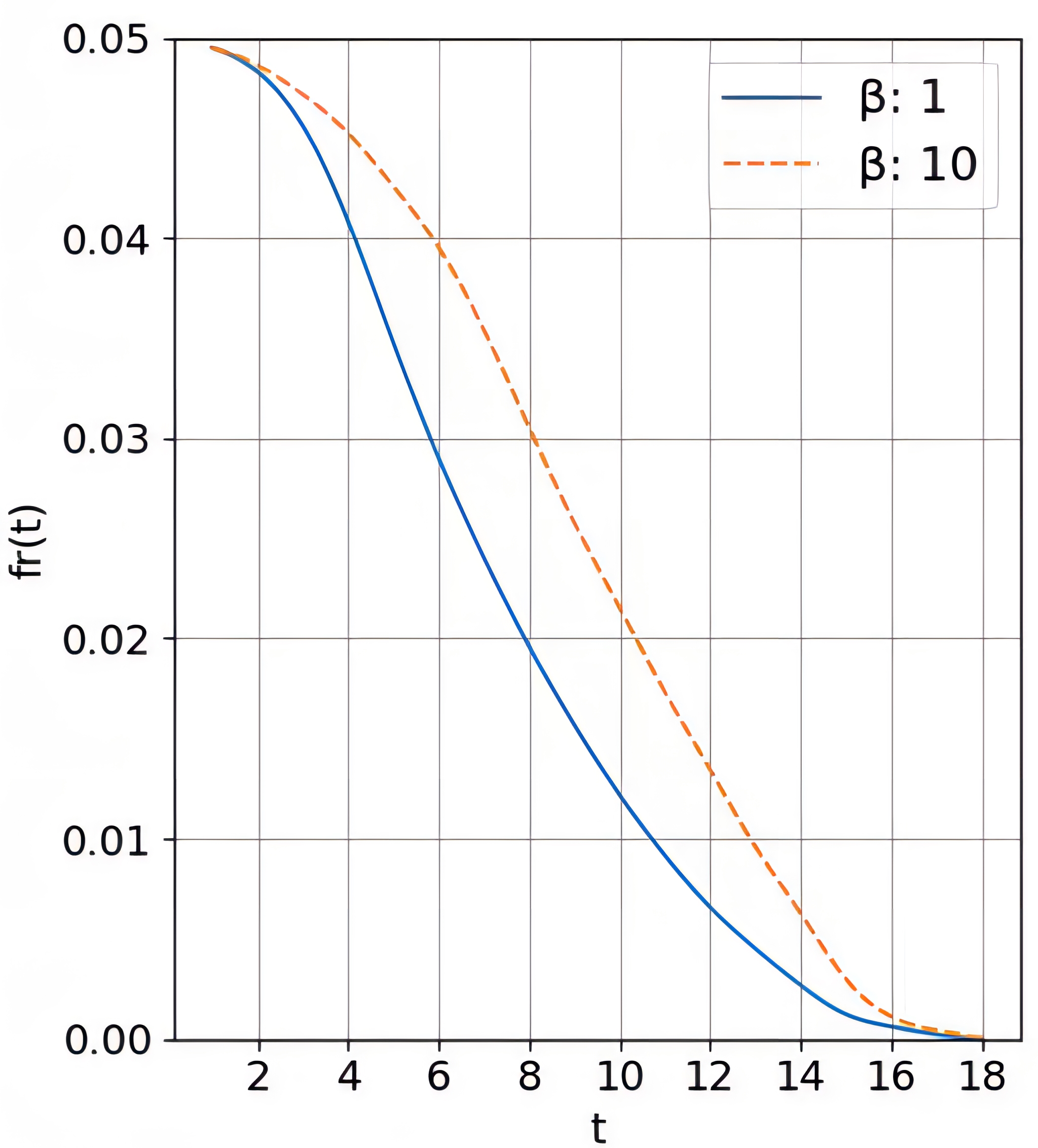}
		\end{minipage}
	}%
	\centering
	\caption{The forking probability in unit time $fr(t)$ when $\beta=1$ and 10.}
	\label{fig:fr(t)_beta}
\end{figure}
\subsection{Evaluation on the unintentional forking model}\label{ex-B}

\begin{figure}
	\centering
	\subfigure[$\beta=1, N_{total}=145$]{
		\begin{minipage}[t]{0.45\linewidth}
			\centering
			\includegraphics[width=1.68in]{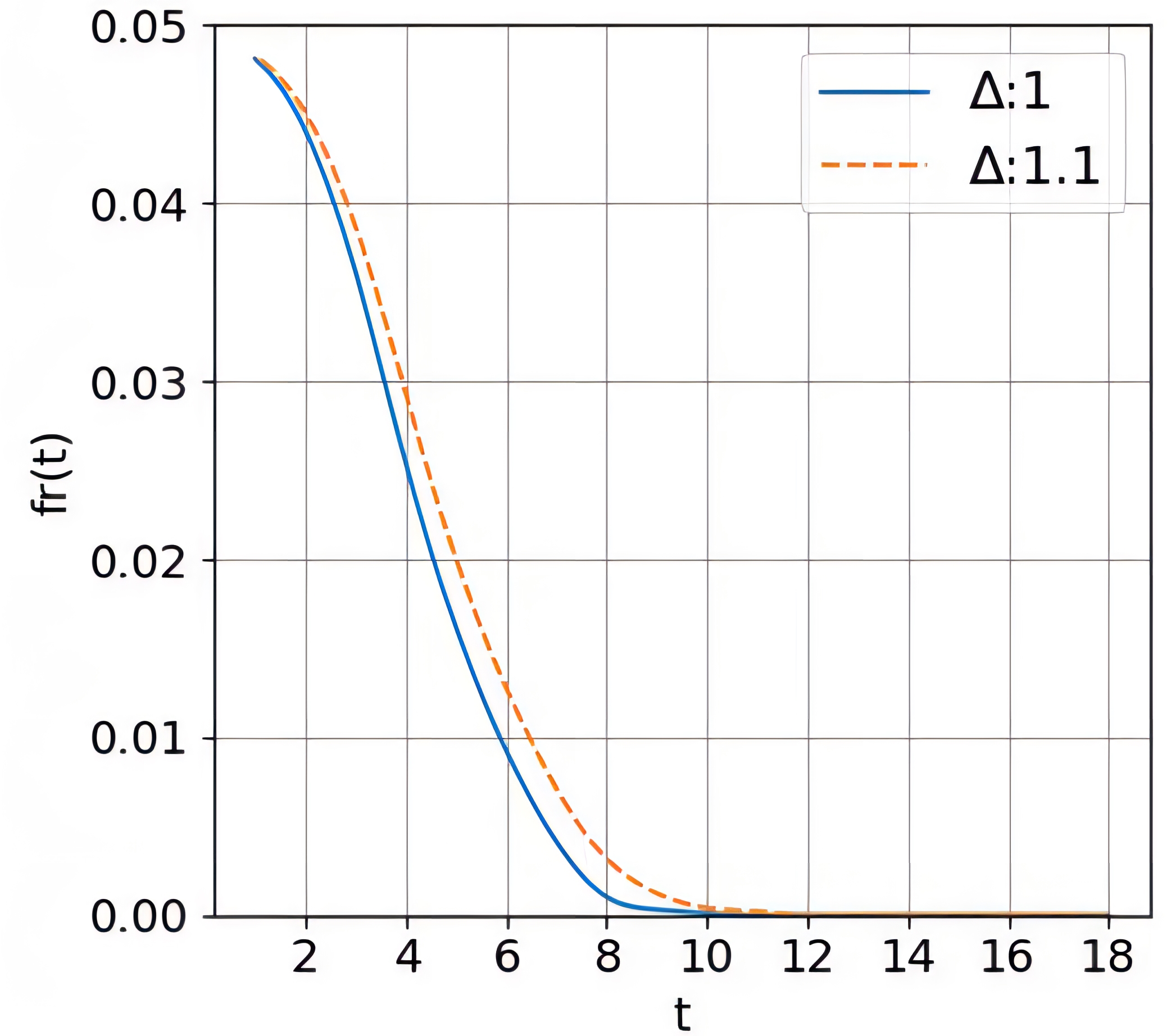}
		\end{minipage}
	}%
	\subfigure[$\beta=1, N_{total}=545$]{
		\begin{minipage}[t]{0.45\linewidth}
			\centering
			\includegraphics[width=1.68in]{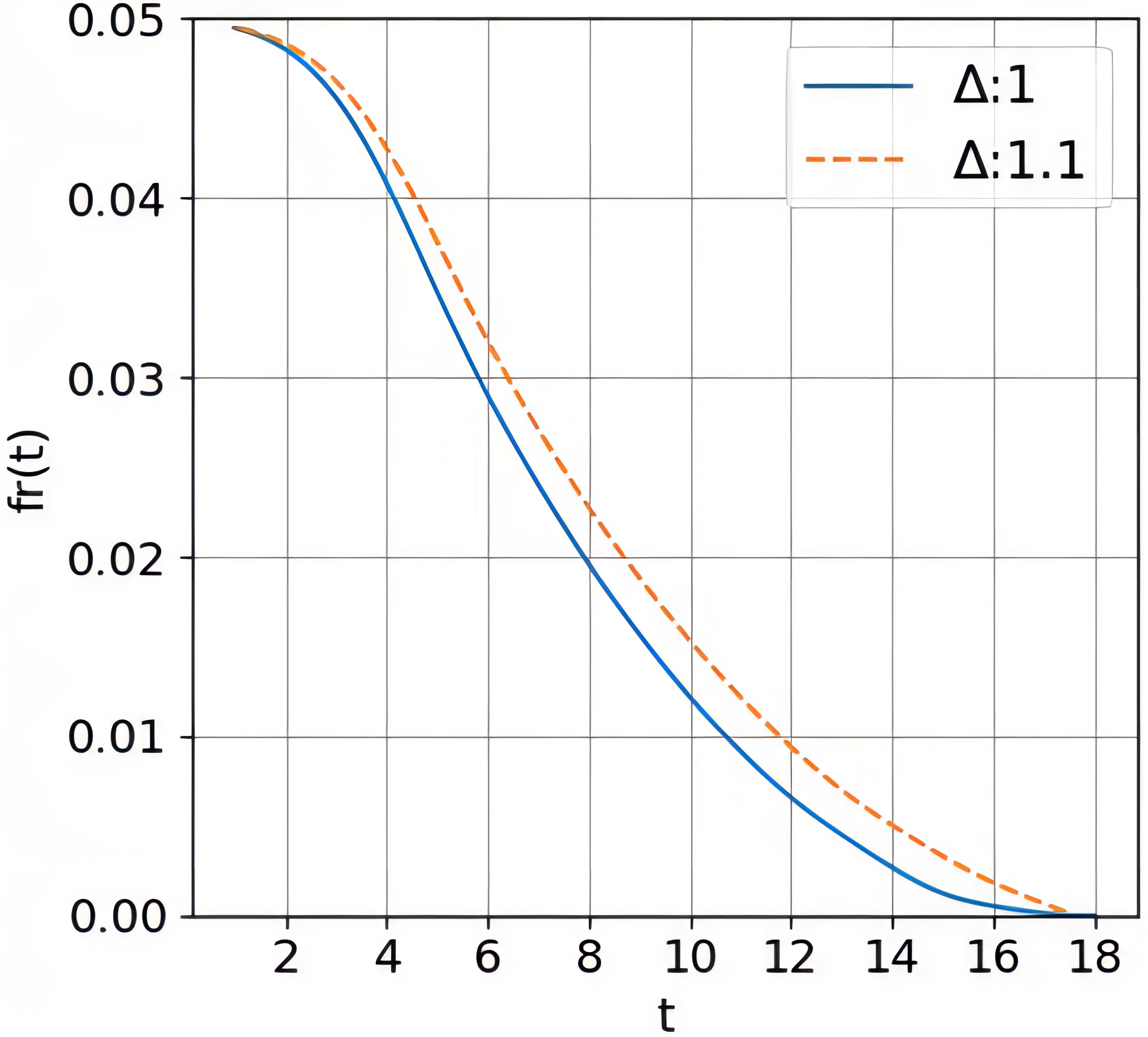}
		\end{minipage}
	}%

	\subfigure[$\beta=10, N_{total}=145$]{
		\begin{minipage}[t]{0.45\linewidth}
			\centering
			\includegraphics[width=1.68in]{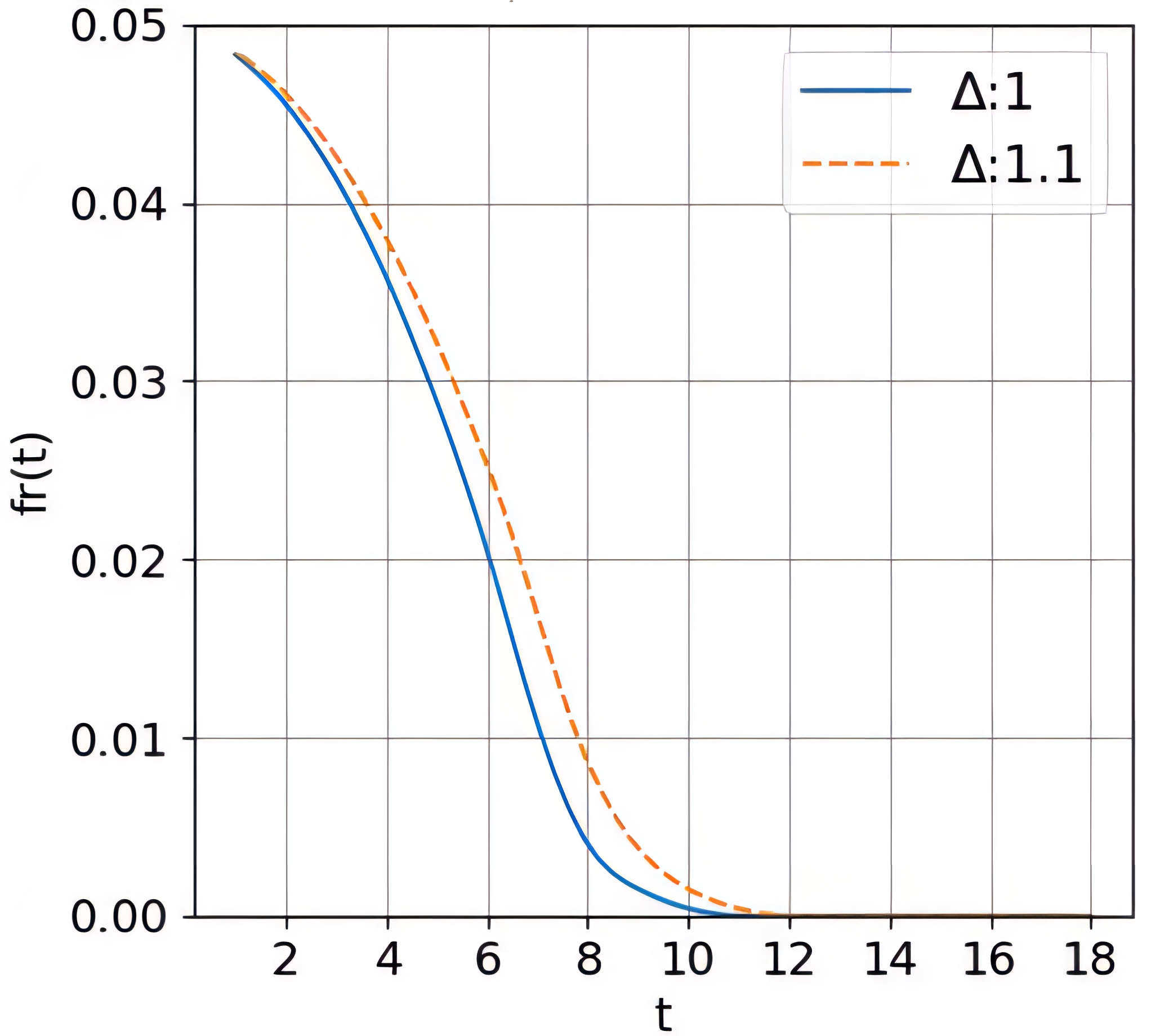}
		\end{minipage}
	}%
	\subfigure[$\beta=10, N_{total}=545$]{
		\begin{minipage}[t]{0.45\linewidth}
			\centering
			\includegraphics[width=1.68in]{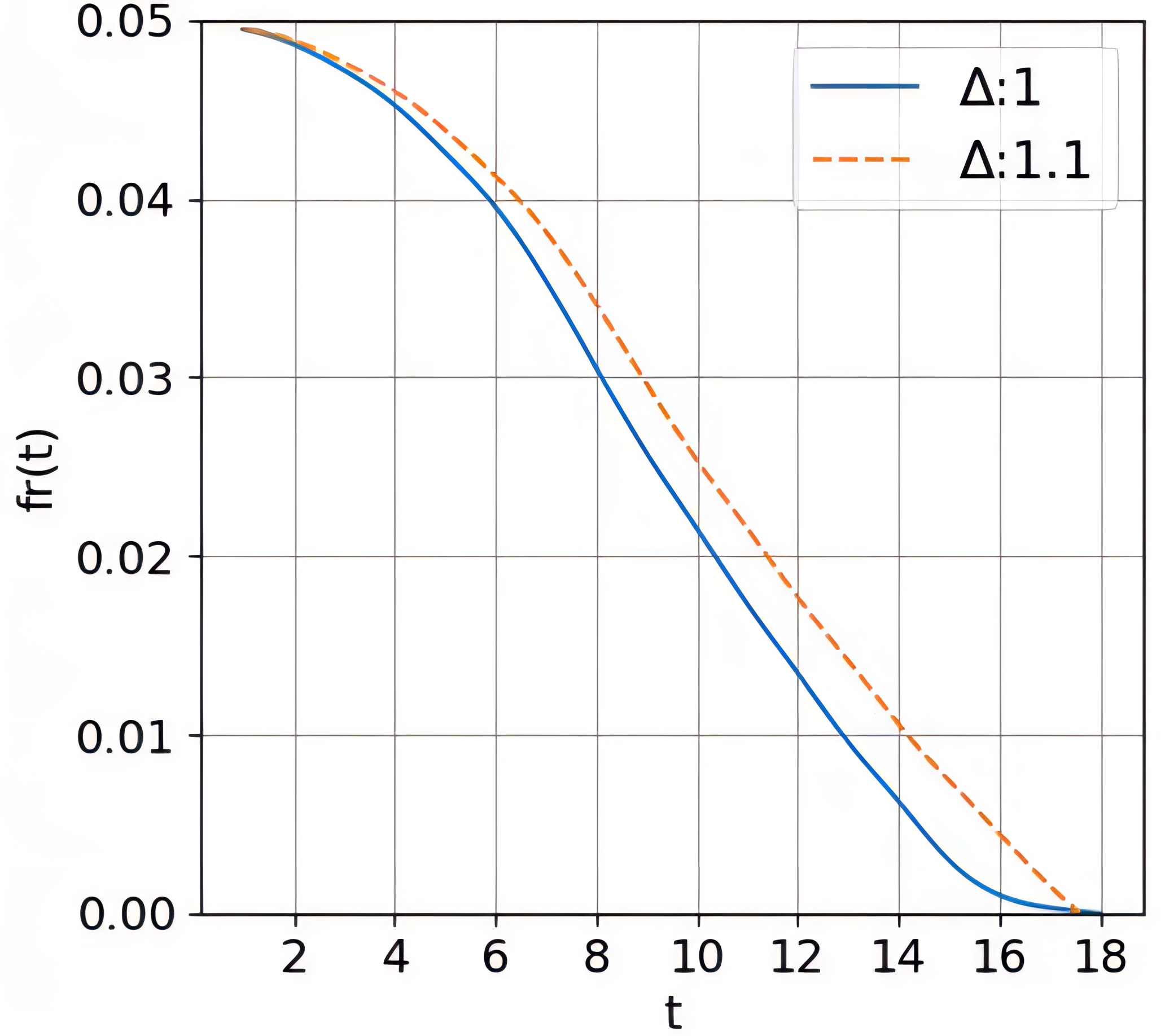}
		\end{minipage}
	}%
	\centering
	\caption{The forking probability in unit time $fr(t)$ when $\Delta=1$ and 1.1.}
	\label{fig:fr(t)_delay}
\end{figure}
After testifying the validity of the main chain propagation model $I(t)$, we proceed to evaluate the performance of forking probabilities $fr(t)$ and $FR(t)$ in unintentional forking. Besides the above described experimental settings, we set the computing power of each node equally and bring in the transmission delay ratio $\Delta$ in which $\Delta=1$ means the initially set $\Delta_s=1$ and $\Delta_l=1.5$, and $\Delta=1.1$ represents $\Delta_s'=1.1\Delta_s$ and $\Delta_l'=1.1\Delta_l$.

We plot the evolutions of unintentional forking probability in unit time $fr(t)$ in Figs. \ref{fig:fr(t)_beta} and \ref{fig:fr(t)_delay}. Specifically, Fig. \ref{fig:fr(t)_beta} displays the negative relationship between $fr(t)$ and $t$, when $\beta=1$ and 10 and $N_{total}=145$ and 545. That is, the forking phenomenon can be hindered probabilistically if the main chain is spread long enough. In addition, the fact that the orange lines lie above the blue ones shows that a lower $\beta$ will lead to a smaller forking probability, corroborating Corollary \ref{c1} as a result. Fig. \ref{fig:fr(t)_delay} exhibits the effect of transmission delays on $fr(t)$, which varies with the propagation time $t$ on the difference of $\beta$ and $N_{total}$. The subfigures in Fig. \ref{fig:fr(t)_delay} suggest that higher delays will trigger forking with a higher probability as we can inspect that the orange lines locate above the blue ones. This proves what we state in Corollary \ref{c2}. 
\begin{figure}
	\centering
	\subfigure[$N_{total}=145$]{
		\begin{minipage}[t]{0.45\linewidth}
			\centering
			\includegraphics[width=1.65in]{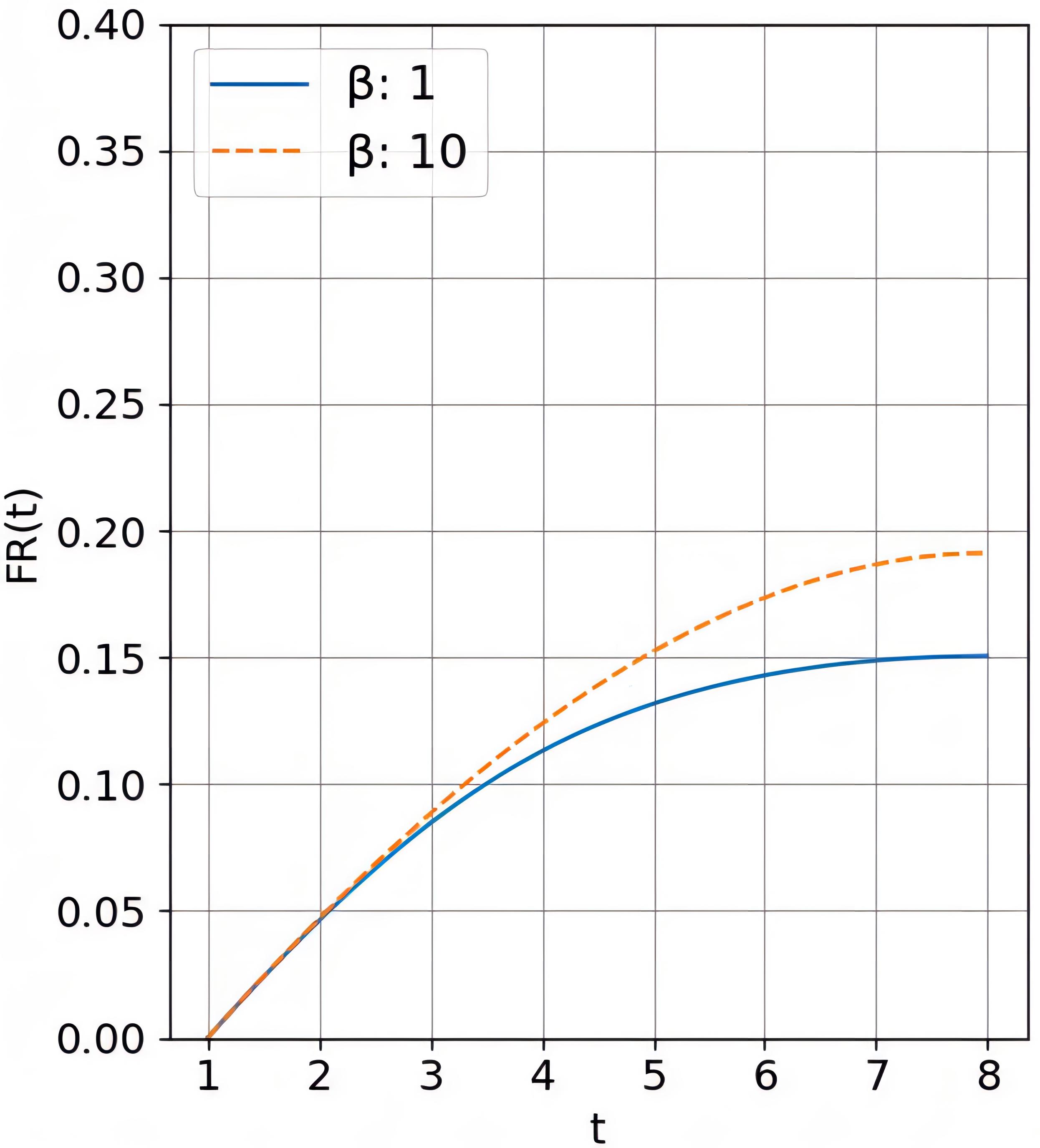}
		\end{minipage}
	}%
	\subfigure[$N_{total}=545$]{
		\begin{minipage}[t]{0.45\linewidth}
			\centering
			\includegraphics[width=1.65in]{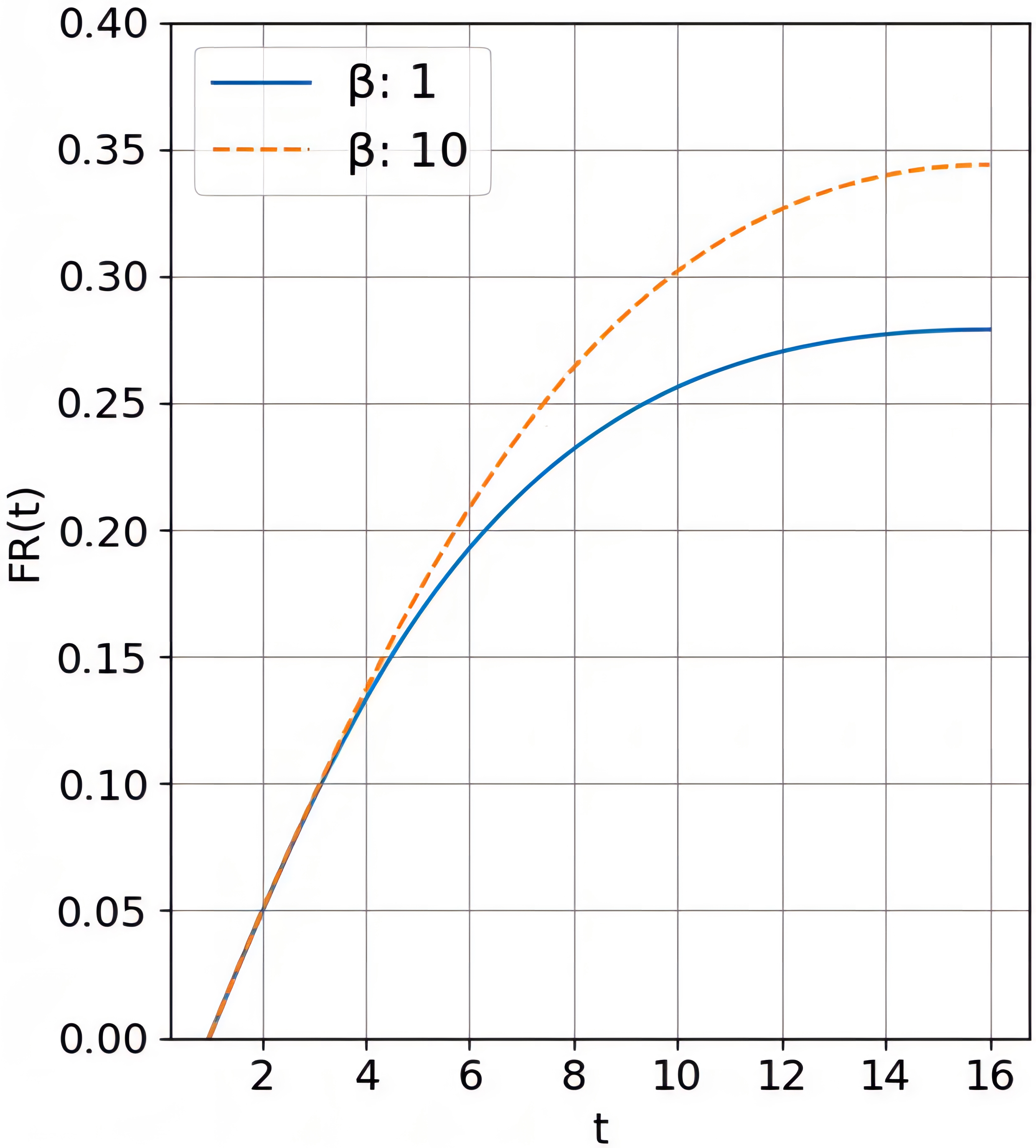}
		\end{minipage}
	}%
	\centering
	\caption{The forking probability over a period of time $FR(t)$ when $\beta=1$ and 10.}
	\label{fig:FR(t)_beta}
\end{figure}

\begin{figure}
	\centering
	\subfigure[$\beta=1, N_{total}=145$]{
		\begin{minipage}[t]{0.45\linewidth}
			\centering
			\includegraphics[width=1.68in]{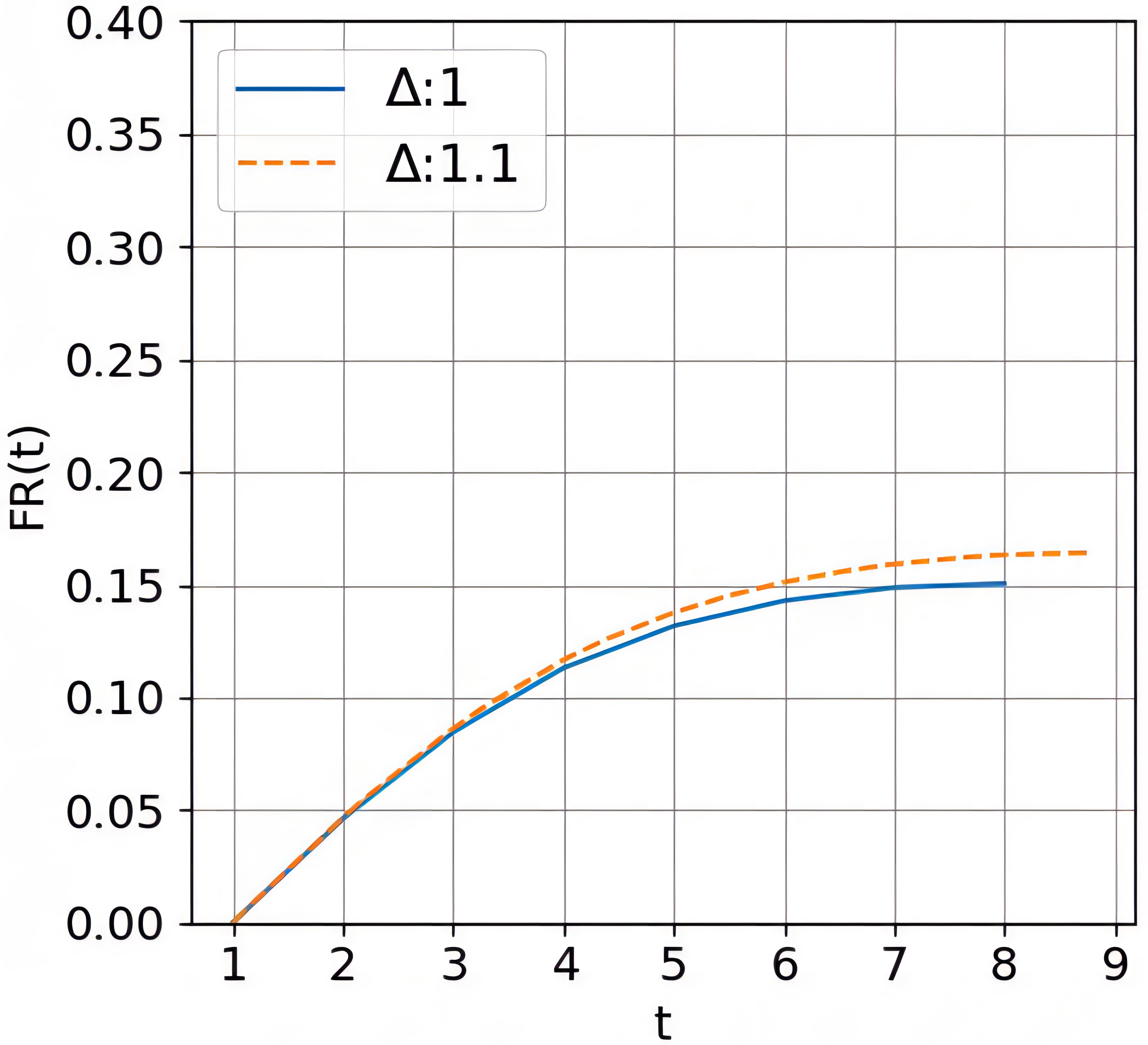}
		\end{minipage}
	}%
	\subfigure[$\beta=1, N_{total}=545$]{
		\begin{minipage}[t]{0.45\linewidth}
			\centering
			\includegraphics[width=1.68in]{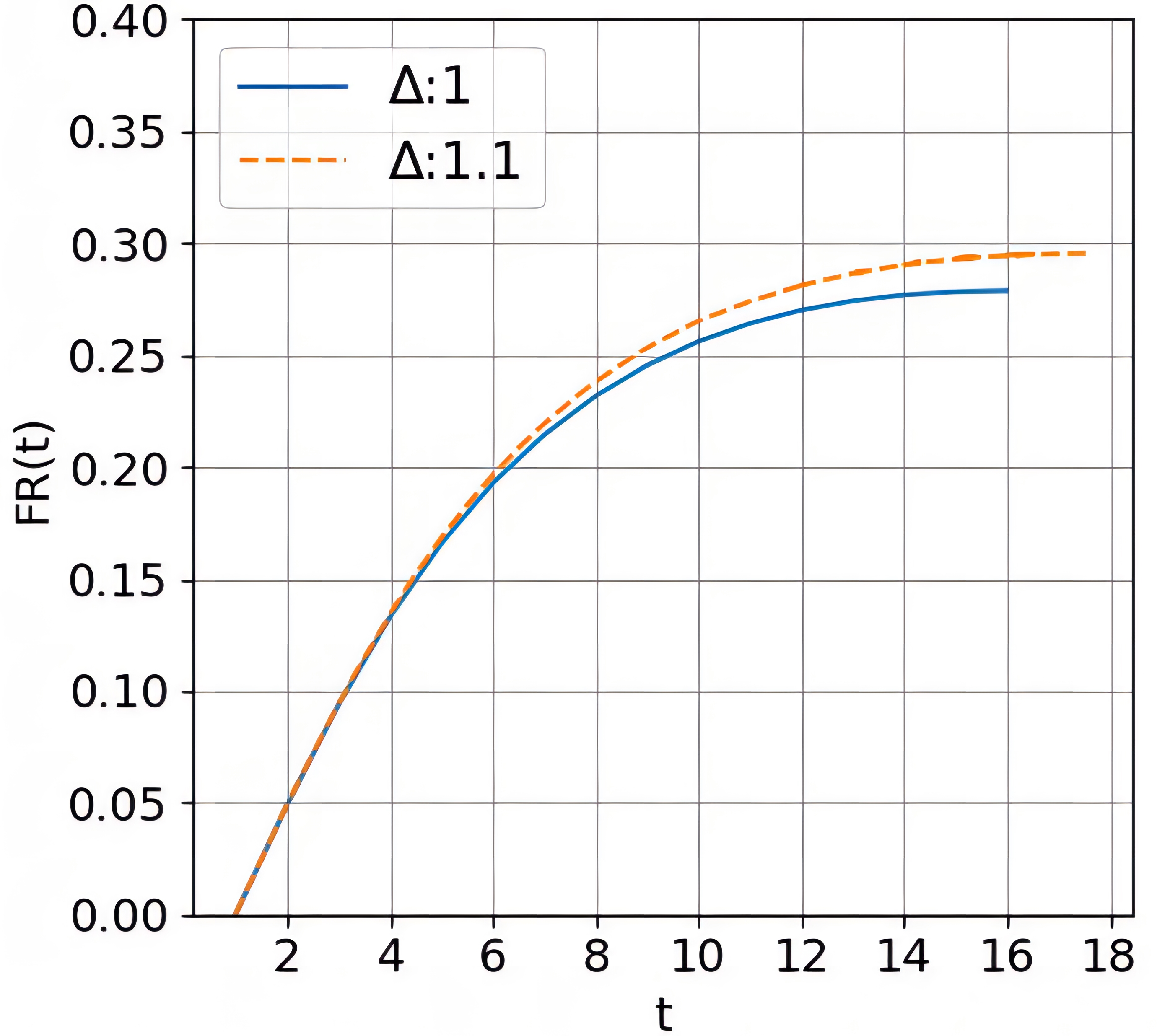}
		\end{minipage}
	}%

	\subfigure[$\beta=10, N_{total}=145$]{
		\begin{minipage}[t]{0.45\linewidth}
			\centering
			\includegraphics[width=1.68in]{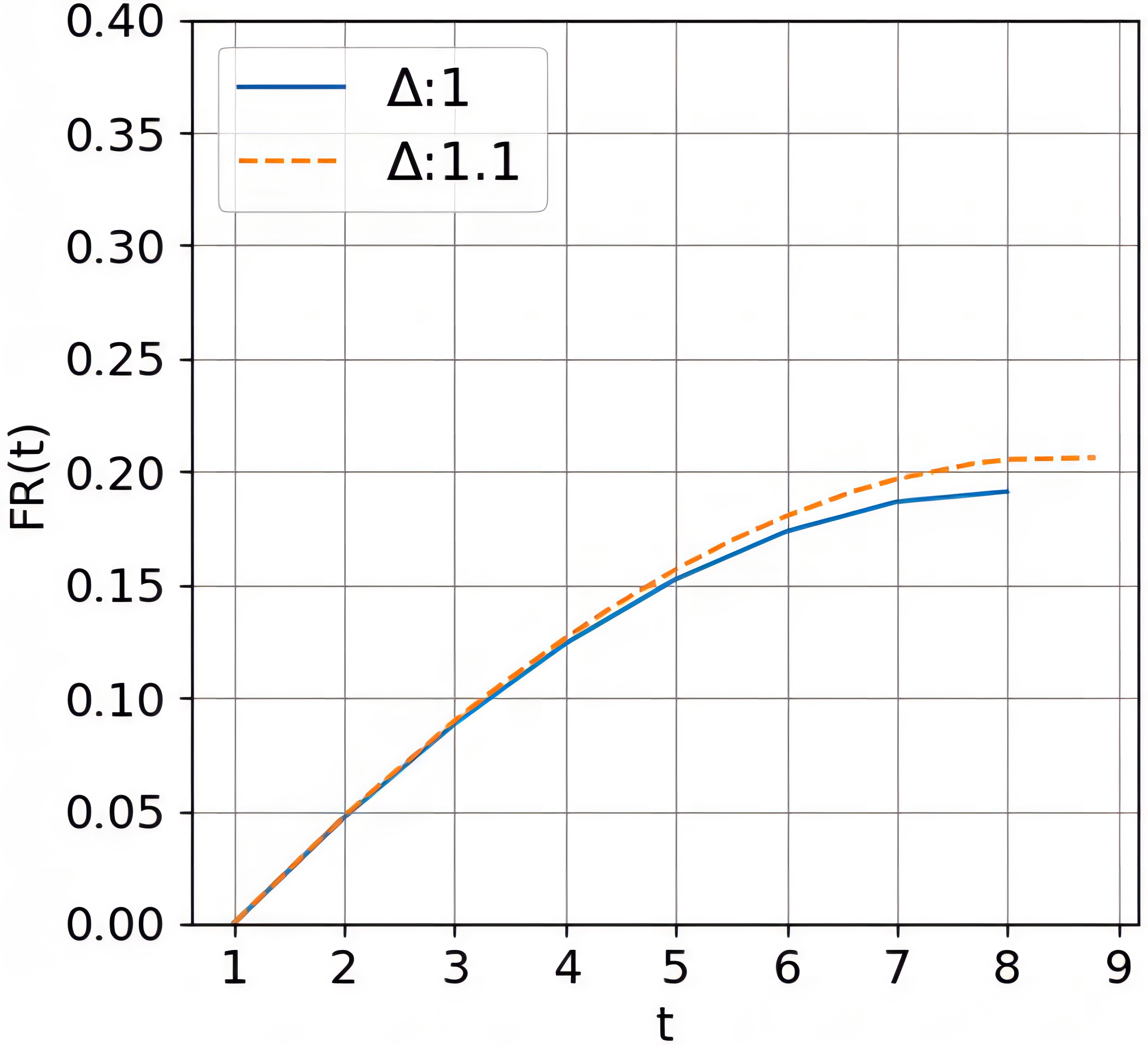}
		\end{minipage}
	}%
	\subfigure[$\beta=10, N_{total}=545$]{
		\begin{minipage}[t]{0.45\linewidth}
			\centering
			\includegraphics[width=1.68in]{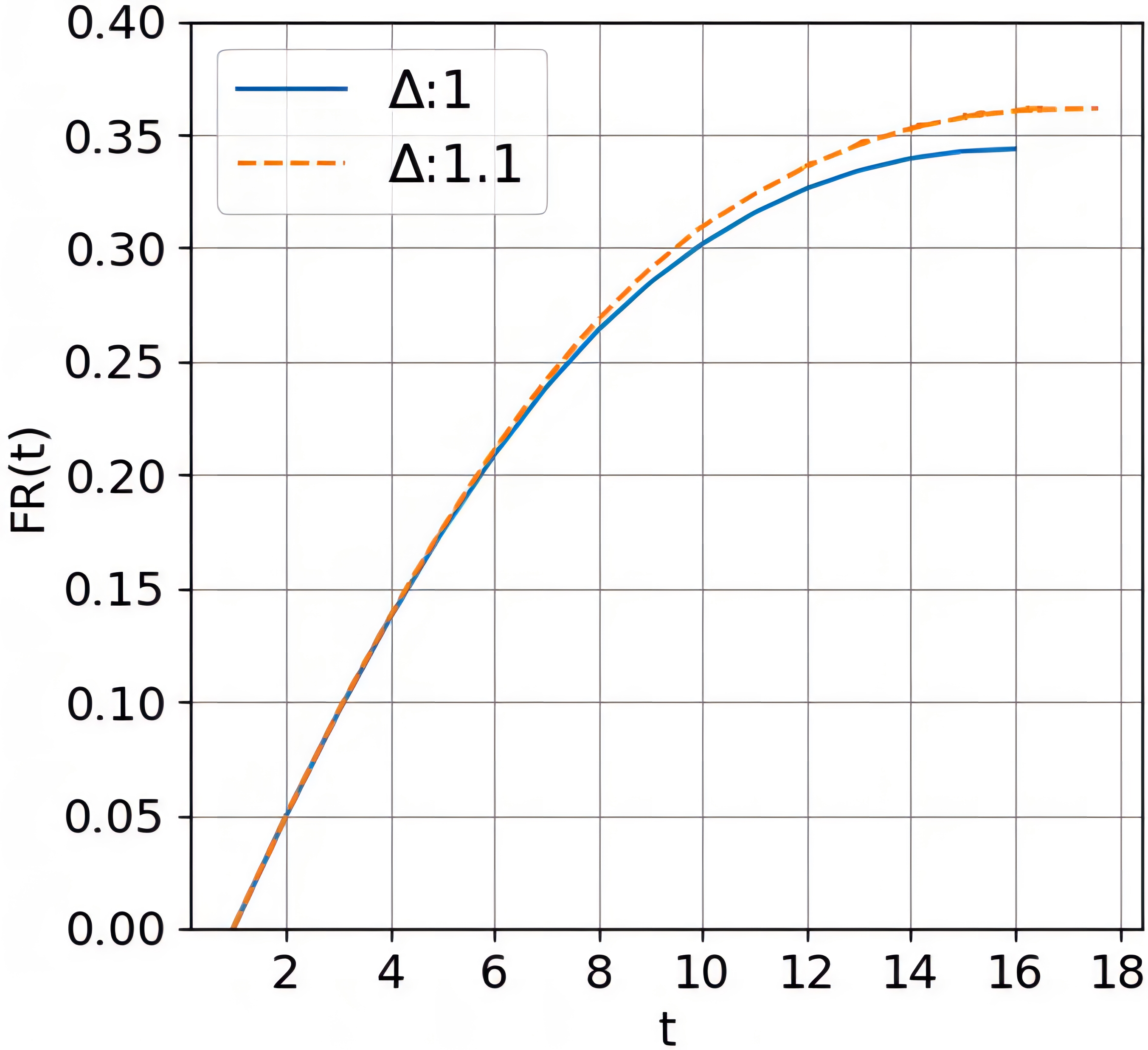}
		\end{minipage}
	}%
	\centering
	\caption{The forking probability over a period of time $FR(t)$ when $\Delta=1$ and 1.1.}
	\label{fig:FR(t)_delay}
\end{figure}

The trends of unintentional forking probability over a period of time, i.e., $FR(t)$, are presented in Figs. \ref{fig:FR(t)_beta}, \ref{fig:FR(t)_delay}, based on which, we can find 1) $FR(t)$ grows as time goes by but with a gradually decreasing rate, which agrees with the proposed analytical model as shown in \eqref{FR(t)}. Theoretically, $FR(t)$ is formalized according to the summation of $S(w)$, that is, the number of suspectable nodes from $w=0$ to $w=t$, which increases with a declining growth rate as more and more nodes accept the main chain. 2) A smaller $\beta$ can incite a lower forking probability as we derived in Corollary \ref{c3} according to Fig. \ref{fig:FR(t)_beta}. 3) Fig. \ref{fig:FR(t)_delay}. testifies the effectiveness of Corollary \ref{c4} in that the forking probability over a period of time $FR(t)$ drops as the decrease of the transmission delays. To conclude, a blockchain overlay network with a lower long-range factor $\beta$, a smaller network size $N_{total}$ and less delays $\Delta_s/\Delta_l$ will endure the least risk of being attacked by unintentional forking.

\subsection{Evaluation on the intentional forking model}\label{ex-C}
\begin{figure}[t]
	\centering
	\subfigure[$N_{total}=145$]{
		\begin{minipage}[t]{0.45\linewidth}
			\centering
			\includegraphics[width=1.65in]{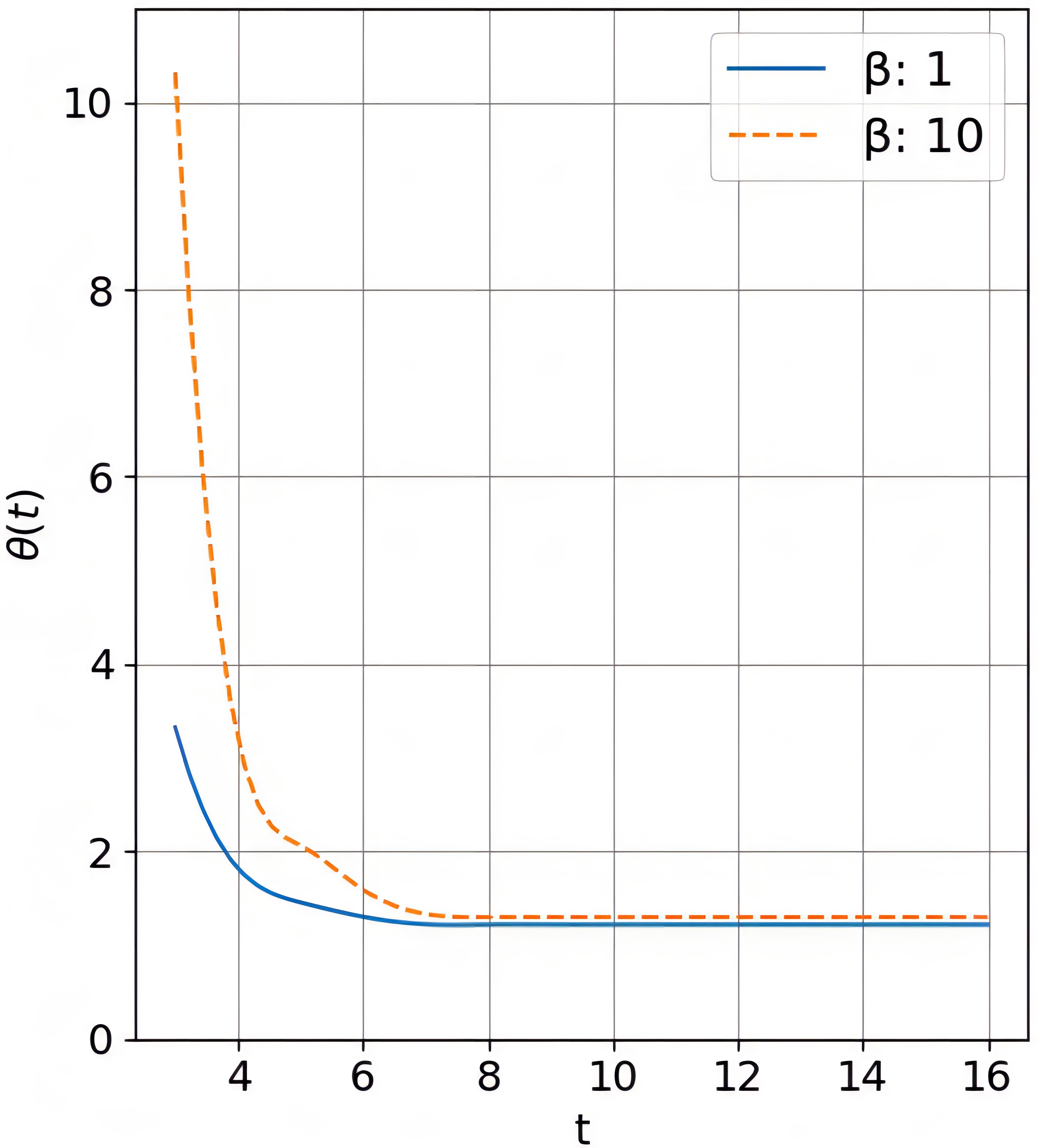}
		\end{minipage}
	}%
	\subfigure[$N_{total}=545$]{
		\begin{minipage}[t]{0.45\linewidth}
			\centering
			\includegraphics[width=1.65in]{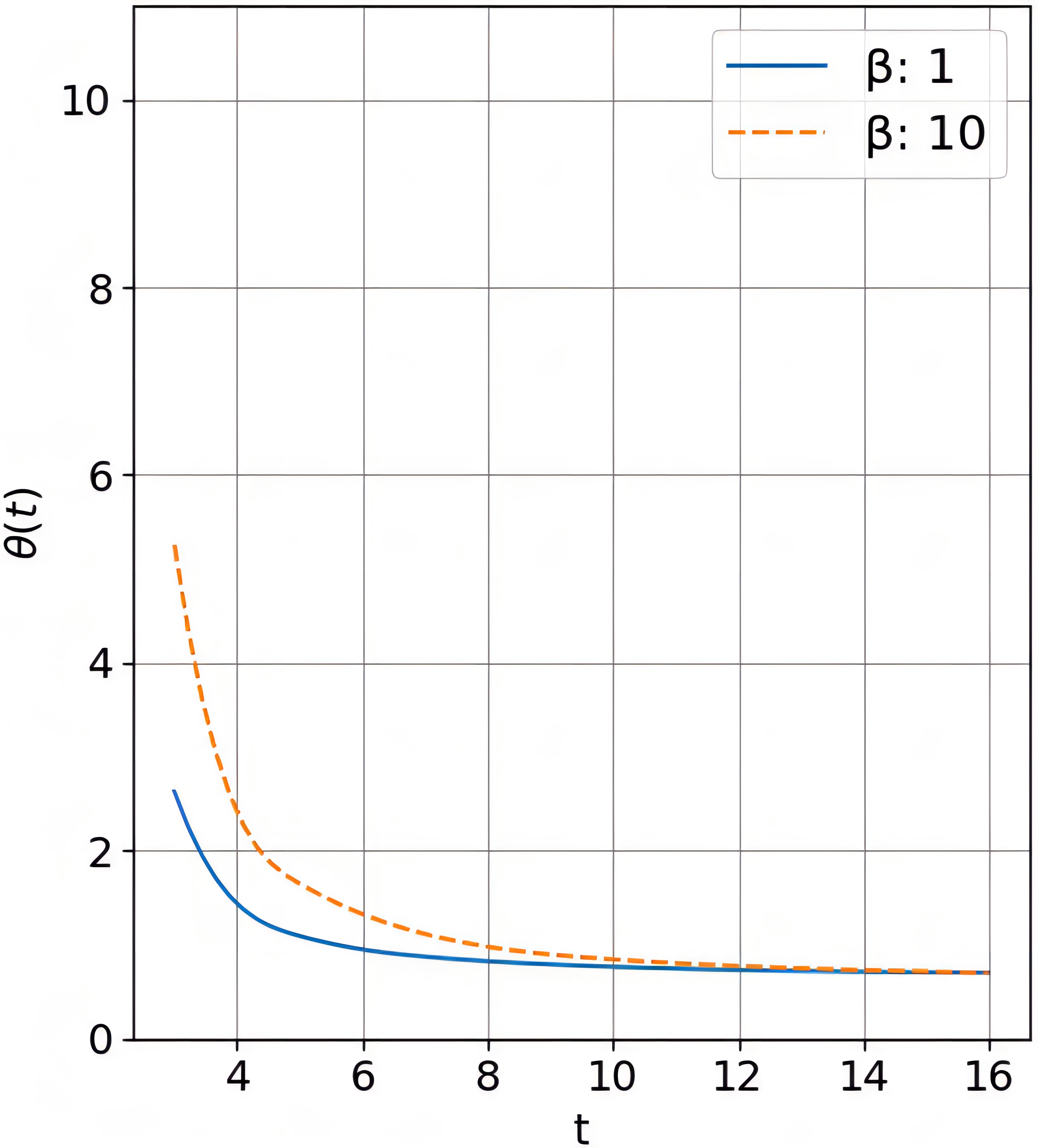}
		\end{minipage}
	}%
	\centering
	\caption{The dynamic robust level with spatiality $\theta(t)$ when $\beta=1$ and 10.}
	\label{fig:robust_beta}
\end{figure}

\begin{figure}[t]
	\centering
	\subfigure[$\beta=1, N_{total}=145$]{
		\begin{minipage}[t]{0.45\linewidth}
			\centering
			\includegraphics[width=1.65in]{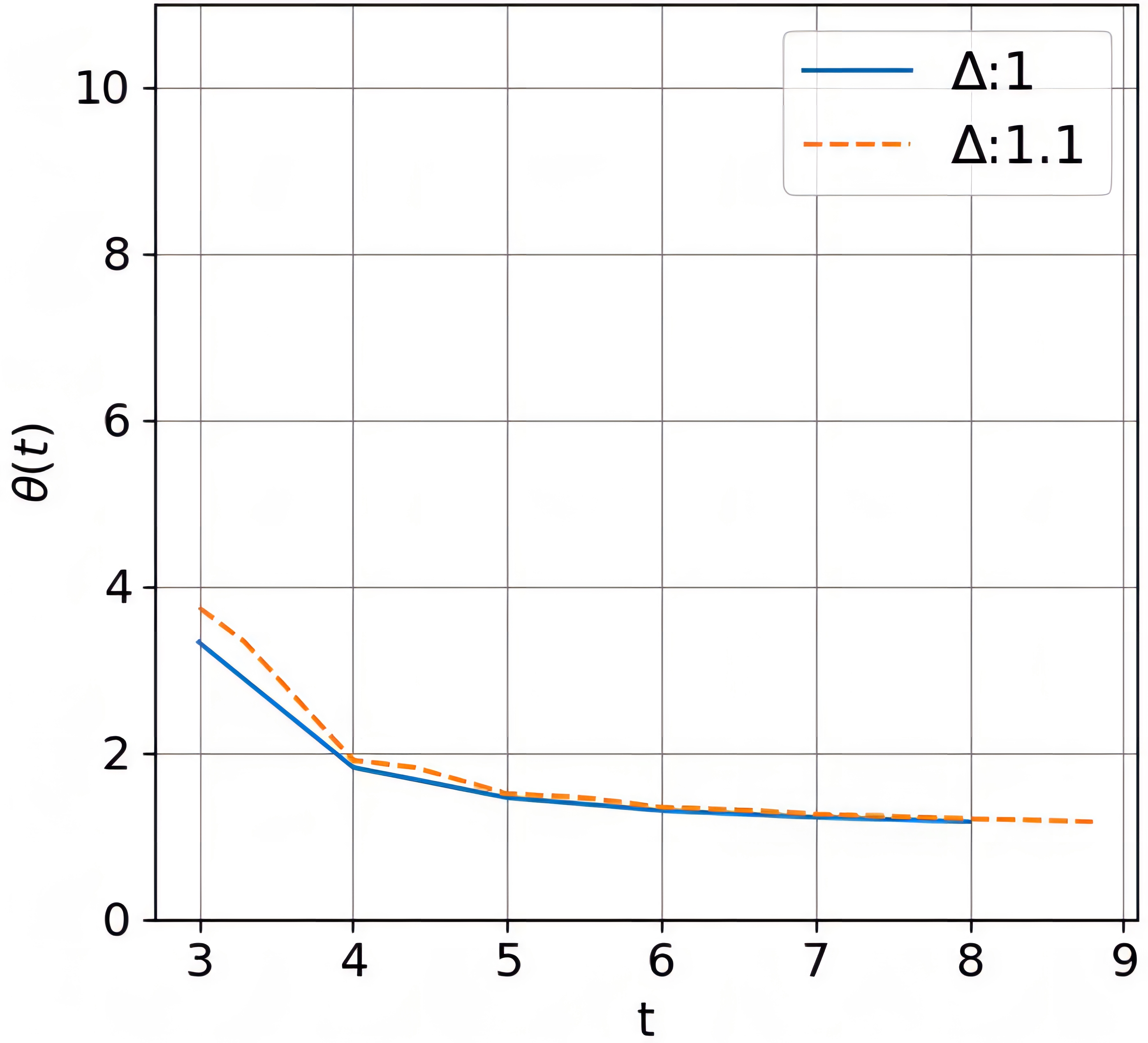}
		\end{minipage}
	}%
	\subfigure[$\beta=1, N_{total}=545$]{
		\begin{minipage}[t]{0.45\linewidth}
			\centering
			\includegraphics[width=1.65in]{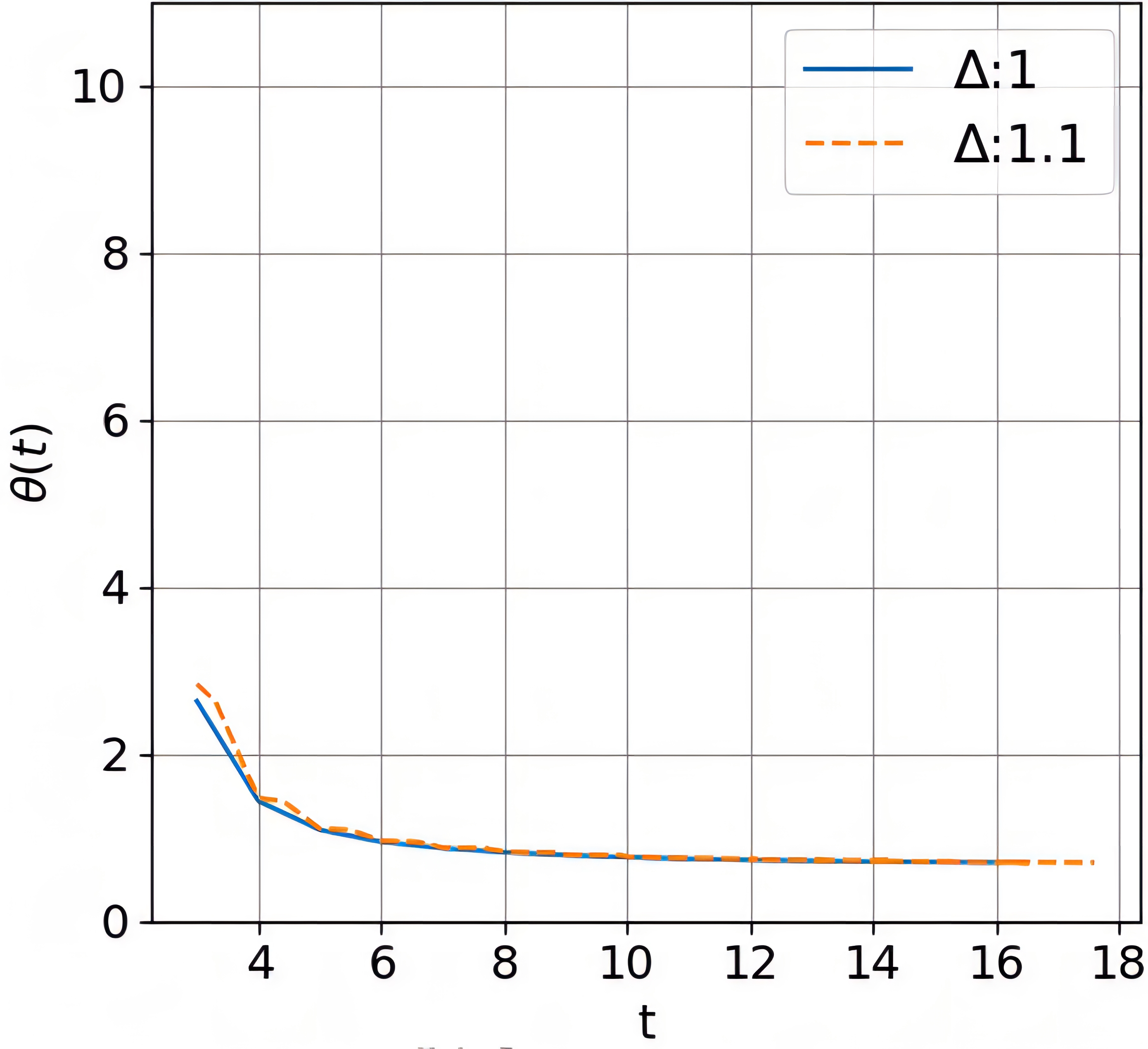}
		\end{minipage}
	}%

	\subfigure[$\beta=10, N_{total}=145$]{
		\begin{minipage}[t]{0.45\linewidth}
			\centering
			\includegraphics[width=1.65in]{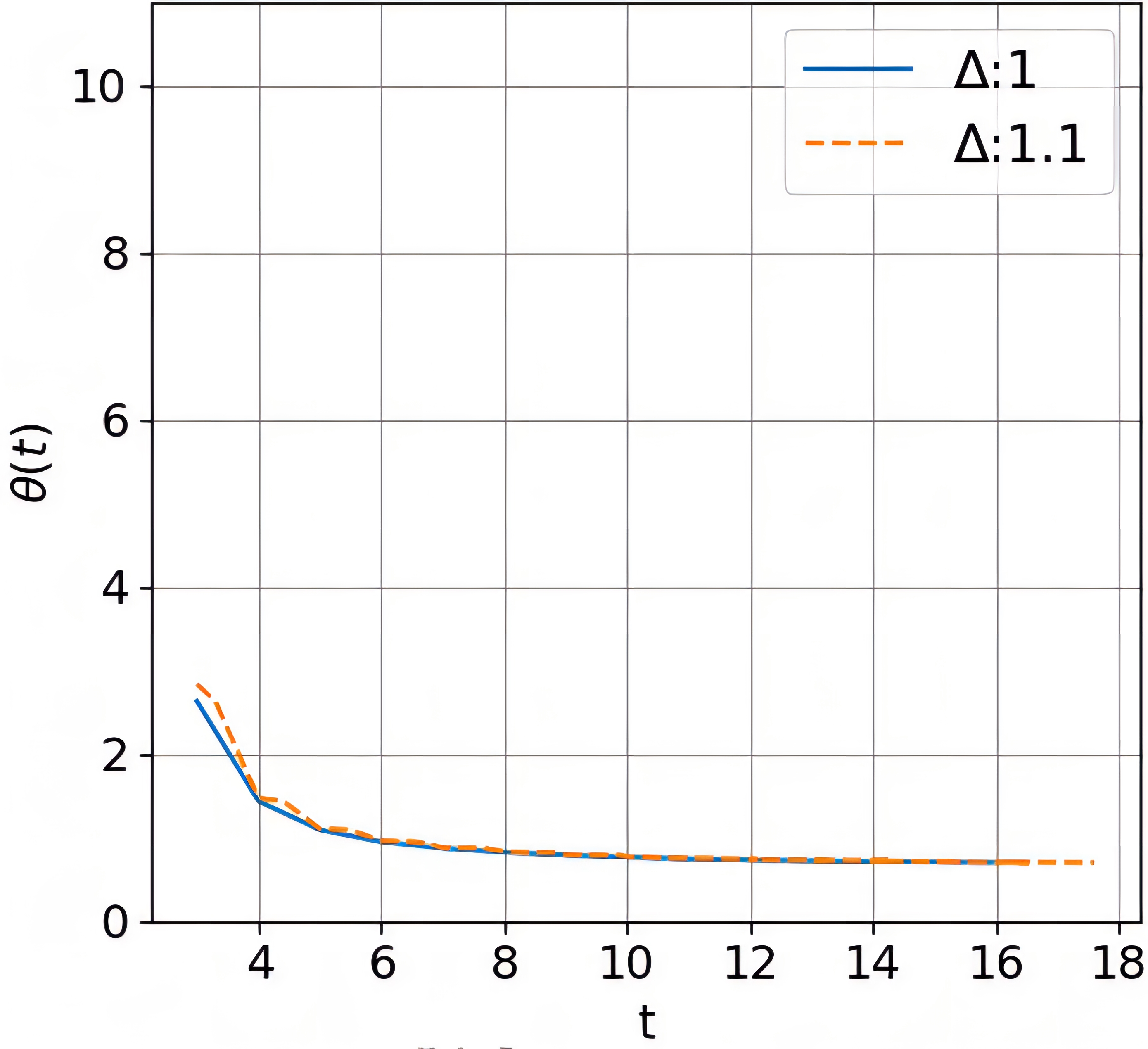}
		\end{minipage}
	}%
	\subfigure[$\beta=10, N_{total}=545$]{
		\begin{minipage}[t]{0.45\linewidth}
			\centering
			\includegraphics[width=1.65in]{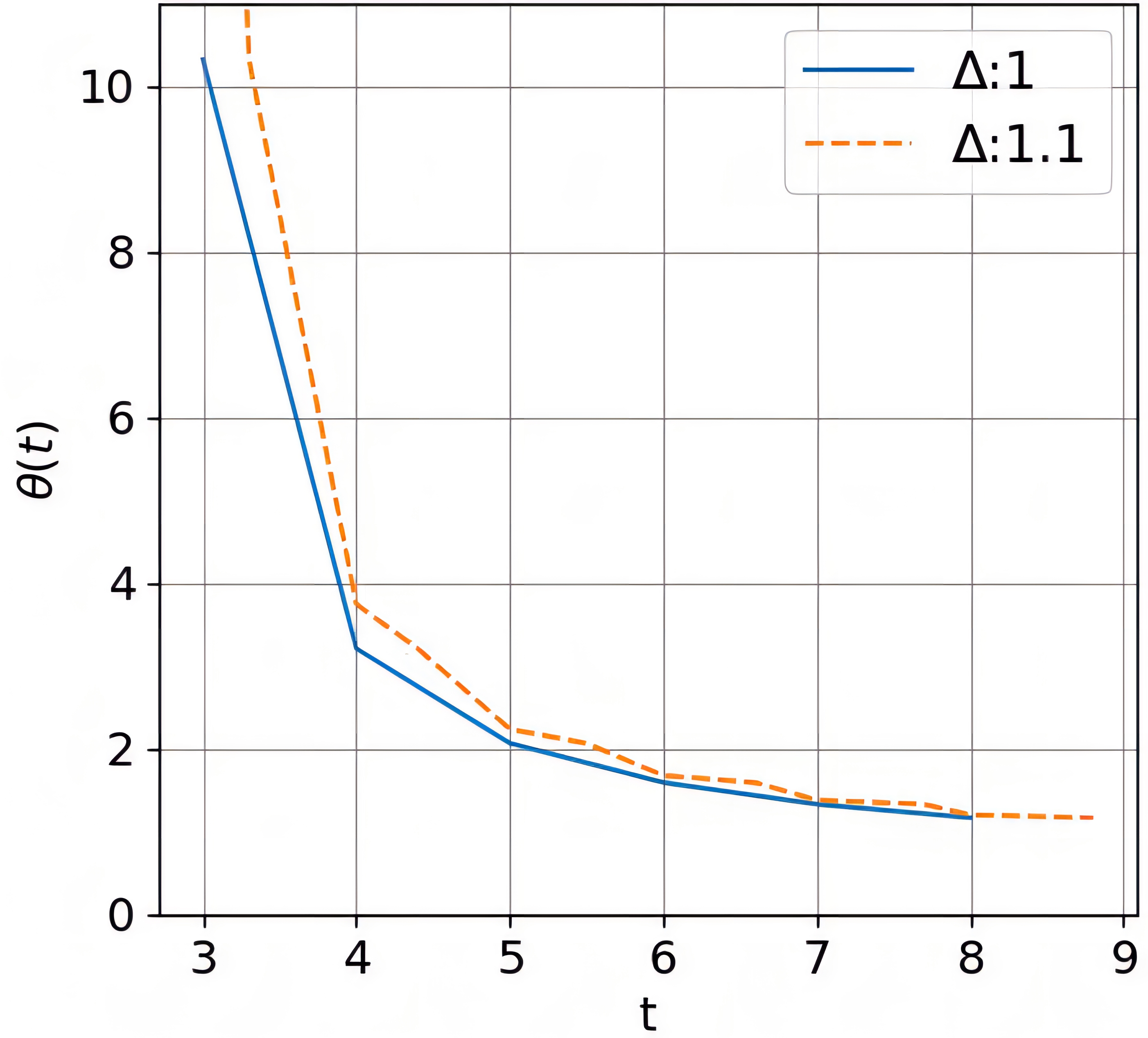}
		\end{minipage}
	}%
	\centering
	\caption{The dynamic robust level with spatiality $\theta(t)$ when $\Delta=1$ and 1.1.}
	\label{fig:robust_delay}
\end{figure}

In this section, we evaluate the dynamic robust level with spatiality, i.e., $\theta(t)$, for the intentional forking scenario. Assume there are $e=20$ malicious nodes, where the threshold $\epsilon$ is preset as 0.1. The results are demonstrated in Figs. \ref{fig:robust_beta} and \ref{fig:robust_delay}. Fig. \ref{fig:robust_beta} reports the relationship between $\theta(t)$ and $t$, from which we can conclude that as time goes by, the robust level will increase since the decrease of $\theta(t)$ denotes less block difference is required for hindering forking occurrence. Besides, it is testified that a smaller $\beta$ can lead to higher $\theta(t)$, as the blue lines are always below the orange ones. This justifies the validity of Corollary \ref{c5}. Additionally, Fig. \ref{fig:robust_delay} shows $\theta(t)$ varies with $t$ when different $\Delta_s$ and $\Delta_l$ are adopted, where we can find lower delays can help promote the robust level as clarified in Corollary \ref{c6}.

\section{Conclusion}\label{sec:conclusion}
In this paper, we carry out the first active defense analysis of blockchain forking from the spatial-temporal dimension. To begin with, we characterize the topology of blockchain network as a two-dimensional grid with distinguished short-/long- range links. Based on this, the concepts of {\it layer} and {\it expansivity} are defined to respectively capture the ripple-effect information propagation process and connection heterogeneity of blockchain. In addition, we propose the term of {\it activation time} to denote the time when the nodes in each layer first accept the main chain, in light of which, we finally realize the {\it activation degree} to express the main chain propagation model. In doing so, the forking probability and robust level in unintentional and intentional forking are modeled and inspected detailedly. Through our analyses, we conclude that 1) dwindling the transmission delays of the short-/long- range links can hinder forking and 2) positively reshaping the overlay network through shrinking the long-range factor $\beta$ can resolve the forking phenomenon from the root. This observation is so valuable that it delivers the first forking defense mechanism proactively and operatively. Extensive theoretical deductions and simulations are conducted to verify the effectiveness of our analysis.

\section*{Acknowledgment}
This work has been supported by National Key R\&D Program of China (No. 2019YFB2102600), National Natural Science Foundation of China (No. 61772044), the International Joint Research Project of Faculty of Education, Beijing Normal University, and Engineering Research Center of Intelligent Technology and Educational Application, Ministry of Education.

\bibliographystyle{IEEEtran}
\bibliography{reference}

\ifCLASSOPTIONcaptionsoff
  \newpage
\fi

\end{document}